\documentclass[11pt]{article}
\usepackage{amsmath, amsthm, amssymb, bbm, setspace,bigints}
\usepackage[margin=1 in]{geometry}
\usepackage{graphicx}
\usepackage{caption}
\usepackage{subcaption}
\usepackage[toc,page]{appendix}
\usepackage{pdfpages}
\usepackage{epstopdf}
\usepackage{booktabs}
\usepackage{multirow}
\usepackage{algorithmic}
\usepackage{algorithm}

\usepackage{natbib}
\usepackage[OT1]{fontenc}
\usepackage[utf8]{inputenc}
\usepackage{authblk}

\usepackage[colorlinks,allcolors=blue]{hyperref}
\newcommand{\be} {\begin{eqnarray*}}
\newcommand{\ee} {\end{eqnarray*}}

\newcommand{\argmin}{\mathop{\rm argmin~}}

\def\T{{ \mathrm{\scriptscriptstyle T} }}

\def\mc{\mathcal}
\def\mb{\mathbb}

\allowdisplaybreaks

\title{Signal Adaptive Variable Selector for the Horseshoe Prior}
\author{Pallavi Ray \thanks{pallaviray@stat.tamu.edu} \quad \quad Anirban Bhattacharya \thanks{anirbanb@stat.tamu.edu}}

\date{}    

\affil{\it Department of Statistics, Texas A\&M University, College Station \\ \it 3143 TAMU, TX 77843-3143, USA}

\providecommand{\keywords}[1]{KEYWORDS: \textit{#1}}

\begin{document}
\maketitle

\begin{abstract}
	In this article, we propose a simple method to perform variable selection as a post model-fitting exercise using continuous shrinkage priors such as the popular horseshoe prior. The proposed Signal Adaptive Variable Selector (SAVS) approach post-processes a point estimate such as the posterior mean to group the variables into signals and nulls. The approach is completely automated and does not require specification of any tuning parameters. We carried out a comprehensive simulation study to compare the performance of the proposed SAVS approach to frequentist penalization procedures and Bayesian model selection procedures. SAVS was found to be highly competitive across all the settings considered, and was particularly found to be robust to correlated designs. We also applied SAVS to a genomic dataset with more than 20,000 covariates to illustrate its scalability.
\end{abstract}

\keywords{Adaptive; horseshoe; shrinkage; sparsity; two-group model.}

\section{Introduction}

Continuous shrinkage priors \citep{griffin2010inference, carvalho2010horseshoe, armagan2013generalized, bhattacharya2015dirichlet} expressed as global-local variance mixtures of Gaussians \citep{polson2010shrink} are routinely used in Bayesian analysis of high-dimensional regression problems. Such priors induce ``approximate'' sparsity in the regression coefficients by allowing a subset of them to be heavily shrunk towards zero, thereby providing ``one-group'' alternatives \citep{polson2010shrink} to the classical ``two-group'' discrete mixture priors with a point mass at zero \citep{mitchell1988bayesian,george1993variable}. A subclass of these priors, such as the horseshoe and Dirichlet--Laplace, possess attractive theoretical properties in sparse settings, including minimax optimality and frequentist validity of uncertainty characterization \citep{ghosh2014posterior,bhattacharya2015dirichlet, van2017adaptive, van2017uncertainty}. Moreover, the continuous nature of these ``global-local'' priors allows for block updating of the regression parameters from conditionally conjugate Gaussian distributions within standard Gibbs sampling algorithms, with recent advances in sampling from high-dimensional structured Gaussian distributions \citep{bhattacharya2016fast} and associated MCMC algorithms \citep{johndrow2017scalable} substantially improving the scalability of such methods to high-dimensional problems.

Shrinking rather than selecting is a defining feature of these priors, based partly on the intuition that a subset of the covariates may have small but non-null effects \citep{polson2010shrink}, in addition to the computational tractability discussed above. An immediate consequence, however, is that the posterior draws for the regression parameters are non-sparse with probability one, which doesn't automatically lead to variable selection. \cite{carvalho2010horseshoe} defined a local shrinkage factor with values between zero and one for each variable as an analogue to the classical posterior inclusion probability \citep{barbieri2004optimal}, and proposed thresholding this shrinkage factor to include/exclude variables; however, the choice of the threshold remains an issue in practice. \cite{bhattacharya2015dirichlet} proposed grouping the entries of posterior medians into null and non-null groups using 2-means clustering. While this doesn't require any tuning parameter, this approach faces issues when there are signals of varying strengths. A more nuanced multi-group clustering procedure for variable selection was proposed by \cite{li2017variable}. 

In this article, we propose a simple yet effective scheme to select variables using the popular horseshoe prior of \cite{carvalho2010horseshoe}, although the methodology is broadly generalizable to other shrinkage priors. The proposed Signal Adaptive Variable Selector (SAVS) approach {\em post-processes} a point estimate such as the posterior mean or median, obtained using MCMC or some other deterministic approximation, to group the variables into signals and nulls. The procedure is entirely automatic and does not involve any tuning parameters. Our approach is partly motivated by \cite{hahn2015decoupling} who posed the variable selection problem in terms of minimizing a posterior expected loss. Their selection is based on examination of ``selection summary plots'' while we provide a more objective and automated procedure.  

We carried out a comprehensive simulation study to compare the performance of the proposed SAVS approach to frequentist penalization procedures such as adaptive lasso \citep{zou2006adaptive},  smoothly clipped absolute deviation or SCAD \citep{fan2001variable} and minimax concave penalty or MCP \citep{zhang2010nearly}, as well as state-of-the-art Bayesian model selection procedures \citep{shin2015scalable}. Overall, SAVS was found to be highly competitive across all the settings considered, and was particularly found to be robust to correlated designs. We also apply SAVS to a genomic dataset with more than 20,000 covariates to illustrate its scalability.

\section{Methodology}
Consider the Gaussian linear regression model:  
\begin{align}\label{model}
y = X \beta + \epsilon, \hspace{5mm} \epsilon \sim \mc N(0,\sigma^2 I_n),
\end{align}
where $X \in \mb{R}^{n \times p}$ is an $n \times p$ matrix of covariates, with the number of variables $p$ potentially much larger than the sample size $n$. We shall work with the horseshoe prior of \cite{carvalho2010horseshoe}, adopting its usage in the high-dimensional regression context from \cite{bhattacharya2016fast}, 
\begin{align*}
\beta_j \mid \lambda_j, \tau \stackrel{\text{ind.}}\sim \mc N(0, \sigma^2 \lambda_j^2 \tau^2), \ \lambda_j \stackrel{\text{ind.}}\sim \mb{C}_{+}(0,1), \ \tau \sim \mb{C}_{+}(0,1),
\end{align*}
where $\mb{C}_{+}(0,1)$ denotes the half-Cauchy distribution with density proportional to $(1+x^2)^{-1} \mathbbm{1}_{(0, \infty)}(x)$. 

\begin{figure}
	\centering
	\includegraphics[width=\textwidth]{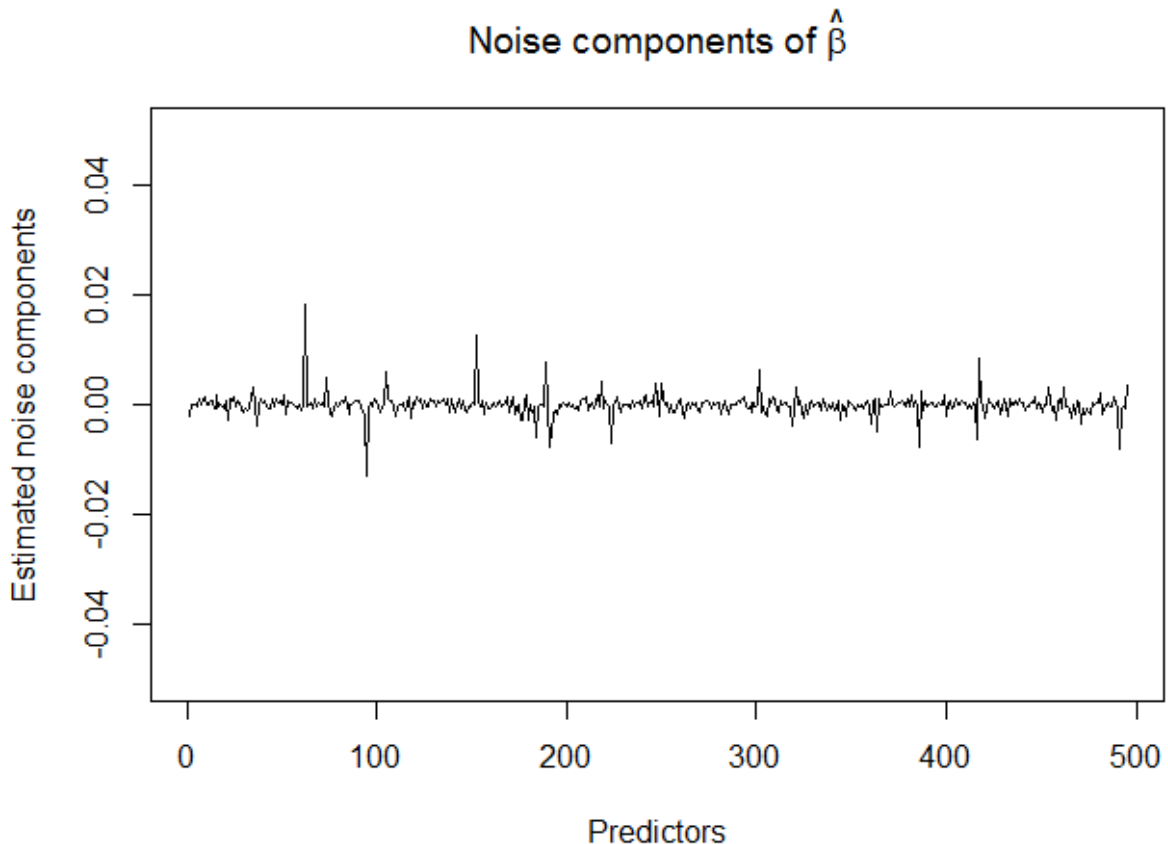}
	\caption{Illustration of the shrinkage performed by the horseshoe prior. The null components of the posterior mean $\widehat{\beta}$ plotted against variable index in a simulation study with $p = 500$, $n = 200$, $\sigma = 1$ and orthogonal design.}\label{noise_comp}
\end{figure}

Let $\hat{\beta} = \displaystyle{\int} \beta \, \Pi_{HS} \big( \beta \mid y \big) d \beta$ be the posterior mean of $\beta$; we throughout used the MCMC algorithm developed in \cite{bhattacharya2016fast} to estimate $\widehat{\beta}$, implemented in the \textbf{{\fontfamily{qcr} \selectfont R}} package \textbf{{\fontfamily{qcr} \selectfont horseshoe}}. As discussed already, the posterior mean $\hat{\beta}$ does not contain exact zeros and instead shrinks the noise coefficients towards zero; see, e.g., Figure \ref{noise_comp}. In the following, we describe a procedure to sparsify $\hat{\beta}$ to obtain a sparse estimator $\hat{\beta}^\ast$.

For a real number $a$, let $\textit{sign}(a) \in \{1, -1\}$ denote its sign with $\textit{sign}(a) = 1$ for $a \ge 0$ and $-1$ otherwise. Also, let $a_+ = \max\{a, 0\}$ denote the positive part of $a$. For a vector $x \in \mathbb{R}^d$, we use $\|x\|$ to denote its Euclidean norm. With these ingredients, let, 
\begin{align}\label{coordinate-descent}
\hat{\beta}_j^\ast = \textit{sign} \big(\hat{\beta_j}\big) \|X_j \|^{-2} \Big( |\hat{\beta_j}| \cdot \|X_j \|^2 - \mu_j \Big)_{+} \, \, , \quad j = 1, \ldots, p,
\end{align}
where, $X_j$ is the $j^{th}$ column of X and $\mu_j = 1/|\hat{\beta_j}|^2, \, j = 1, \ldots, p$. We henceforth refer to $\hat{\beta}^\ast$ as the \textit{Signal Adaptive Variable Selector} (\textbf{SAVS}) estimator; a pseudo-code is provided as Algorithm~\ref{algo}. 

\begin{algorithm}
	\caption{SAVS Algorithm:}
	\begin{algorithmic}\label{algo}
		\REQUIRE Posterior mean $\hat{\beta}$ and design matrix X
		\FOR{$j=1$ \TO $p$}
		\STATE $\mu_j =  1/|\hat{\beta_j}|^2$
		\IF {$|\hat{\beta_j}| \cdot \|X_j \|^2 \leq \mu_j$} 
		\STATE $\hat{\beta}_j^\ast = 0$
		\ELSE
		\STATE $\hat{\beta}_j^\ast = \textit{sign} \big(\hat{\beta_j}\big) \|X_j \|^{-2} \Big(|\hat{\beta_j}| \cdot \|X_j \|^2 - \mu_j \Big)$
		\ENDIF 
		\ENDFOR
		\ENSURE A \textit{sparse} estimate $\hat{\beta}^\ast$
	\end{algorithmic}
\end{algorithm}
The SAVS algorithm takes a non-sparse point estimate $\hat{\beta}$ and the design matrix $X$ as input, and returns a sparse 
estimate $\hat{\beta}^\ast$ which can be readily used for variable selection. While we use the horseshoe prior to obtain $\hat{\beta}$, one can use other shrinkage priors from the vast library of such priors available now. Certainly, the quality of  SAVS would depend on the efficacy of the shrinkage performed. Our choice of the horseshoe was motivated by its automated nature, i.e., no tuning parameters, and its impressive performance documented across a wide range of simulation studies \citep{carvalho2010horseshoe, bhattacharya2015dirichlet, bhattacharya2016fast, van2017adaptive}. 

\subsection{Motivation}
We now provide some intuition behind SAVS. Having obtained the shrinkage estimator $\hat{\beta}$, a natural way to obtain a class of sparse estimators is to solve the optimization problem 
\begin{align}\label{eq:adLa}
\hat{\beta}^\ast :\,= \argmin_{\beta} \bigg\{ \frac{1}{2} \|X \hat{\beta} - X \beta\|_2^2 + \sum_{j=1}^{p} \mu_j |\beta_j| \bigg\}, 
\end{align}
which aims to find a sparse value of $\beta$ close to $\hat{\beta}$ in terms of the Euclidean distance between the ``model fit'' $X \hat{\beta}$ and $X \beta$. The variable specific parameters $\mu_j \ge 0$ control the amount of penalization for each variable akin to adaptive lasso. In the specific situation when $n = p$, the design $X$ is orthogonal, and $\mu_j = \mu$ for all $j$, the optimization problem can be exactly solved, with $\hat{\beta}^\ast_j = \textit{sign}(\hat{\beta}_j) \, (|\hat{\beta}_j| - \mu)_+$ the well-known soft-thresholding estimator. More generally, the optimization problem (\ref{eq:adLa}) is a convex problem and can be solved using the \texttt{R} package \texttt{parcor}. However, the choice of the $p$ tuning parameters remains an issue, with cross-validation expensive for large $p$. 

Our contributions towards the development  of the SAVS algorithm are two-fold. First, we exploit the estimator $\hat{\beta}$ to provide default recommendations for the $p$ tuning parameters $\mu_j$, completely avoiding the need to perform cross-validation. The horseshoe mean $\hat{\beta}$ aggressively shrinks the noise components of the true $\beta_0$ towards zero, while retaining the larger signals. We then let $\mu_j = 1/|\hat{\beta_j}|^2$ for $j = 1, \ldots, p$, so that the penalties for the variables are ranked in inverse-squared order of the magnitude of the corresponding coefficient. We have experimented with $\mu_j = 1/|\hat{\beta_j}|^\kappa$ for various values of $\kappa$ and found $\kappa = 2$ to be a reasonable default choice. 

Let $Q(\beta)$ denote the objective function in the right hand side of \eqref{eq:adLa} with the above choices of the $\mu_j$'s. A standard way to optimize $Q(\cdot)$ is the coordinate descent algorithm \citep{friedman2007pathwise}. For a given current state $\tilde{\beta}$, express the objective function in terms of variable $j$ as 
\begin{align*}
\tilde{Q}_j(\beta_j) = \bigg\{ \frac{1}{2} \|X \hat{\beta} - X_{-j} \tilde{\beta}_{-j} - X_j \beta_j \|_2^2 + \sum_{k \neq j} \mu_k |\tilde{\beta}_k| + \mu_j |\beta_j| \bigg\},
\end{align*}
where $X_j$ is the $j^{th}$ column of $X$, $X_{-j}$ is the matrix formed by deleting $j^{th}$ column of $X$ and $\tilde{\beta}_{-j}$ is the $(p-1)$-vector formed by deleting the $j^{th}$ component of $\tilde{\beta}$. Minimizing $\tilde{Q}_j$ over $\beta_j$ now amounts to simply solving a soft-thresholding problem. Thus, the coordinate descent algorithm starts at some initial value $\tilde{\beta}^{(0)}$ and at any given iteration $t$, it makes a pass through the $p$ variables one-at-a-time to obtain the next iterate $\tilde{\beta}^{(t+1)}$ as 
\begin{align}\label{coordinate descent}
\tilde{\beta}_j^{(t+1)} = \frac{1}{X_j^\T X_j} \textit{sign} \big(X_j^\T R_j^{(t)}\big) \big( |X_j^\T R_j^{(t)}| - \mu_j \big)_{+}  \, \, , \, \, j = 1, \ldots, p
\end{align}
where, $R_j^{(t)}$ is partial residual vector due to regression of $X \hat{\beta}$ on $X$ excluding the $j^{th}$ predictor at the $t^{th}$ step, i.e. $R_j^{(t)} = X \hat{\beta} - X_{-j} \tilde{\beta}_{-j}^{(t)}$ , $t \geq 0$.

\begin{figure}
	\centering
	\includegraphics[width=\textwidth, height=0.58\textwidth]{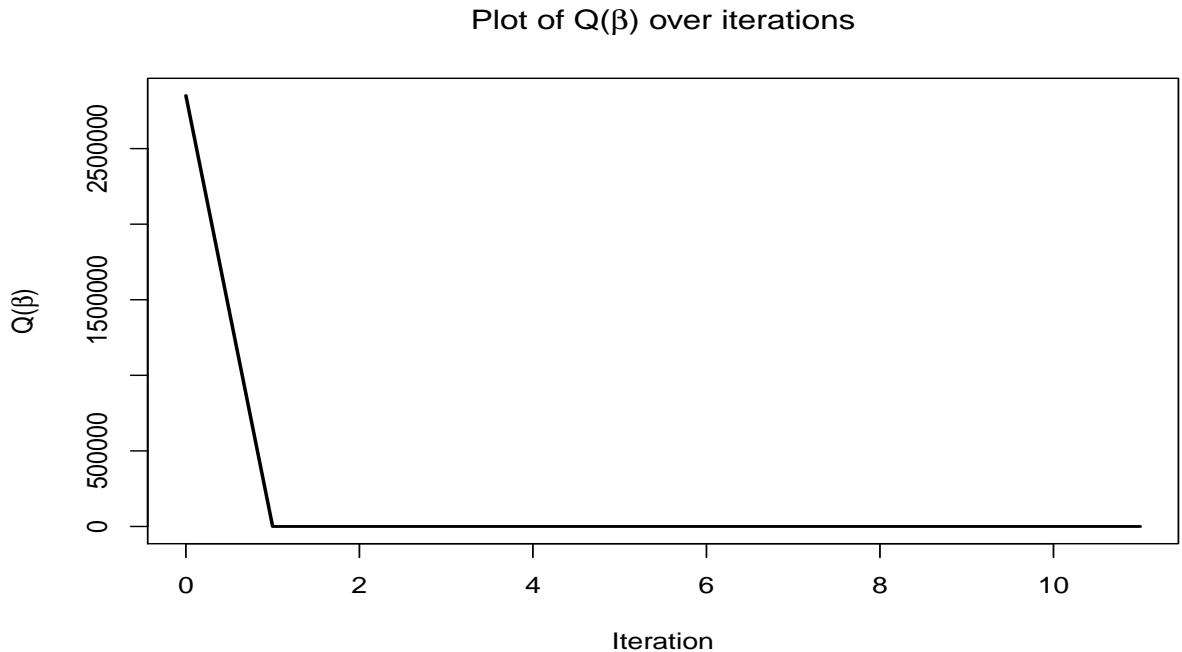}
	\caption{Plot of the objective function $Q(\beta)$ in \eqref{eq:adLa} against iterates of a coordinate descent algorithm. It is evident that convergence takes place after first iteration which is the basis of our early stopping rule. } \label{obj-fun-plot}
\end{figure}

Given the availability of the posterior mean $\hat{\beta}$, it is natural to initialize the algorithm at $\tilde{\beta}^{(0)} = \hat{\beta}$. With this choice, we have noticed that convergence almost always takes place after the first iteration; see Figure \ref{obj-fun-plot} for a representative example. Stopping the algorithm at the first iterate leads to the proposed SAVS algorithm, since $R_j^{(0)} = R_j = \hat{\beta_j} X_j$ and $X_j^\T R_j^{(0)} = X_j^\T R_j = \hat{\beta_j} \|X_j \|^2  \, \, , \, \, j = 1, \ldots, p$. 

When the true $\beta_0$ is sparse, the minimax optimal rate in prediction loss, $\|X \hat{\beta}^\ast - X \beta_0\|^2/n$, is $s_0 \log p/n$ where $s_0 = |S_0|$ and $S_0 : \, = \{\, j : \beta_{0j} \neq 0\}$. Under the additional assumption of coherence conditions for the design matrix, such as the restricted isometry property (RIP; \cite{buhlmann2011statistics}), it can be shown that $\| \hat{\beta}^\ast - \beta_0 \|_1 \lesssim s_0 \sqrt{\log p/n}$. Minimax optimality of the horseshoe in the prediction loss has been established in \cite{chakraborty2016bayesian, van2017uncertainty, van2017adaptive}. Thus, if we additionally assume the RIP condition, then for the horseshoe posterior mean $\hat{\beta}$, we have $|\hat{\beta}_j| = |\beta_{0j}| + \delta_n$ if $j \in S_0$ and $|\hat{\beta_j}| = \delta_n$ if $j \in S_0^C$, where $\delta_n$ is of order $s_0 \sqrt{\log p/n}$. We also assume that, $\|X_j \|^2 \asymp n$. We can rewrite \eqref{coordinate-descent} as:
$$\hat{\beta}_j^\ast = \hat{\beta_j} \, \Bigg\{ 1 - \frac{\mu_j}{|\hat{\beta_j}| \cdot \|X_j \|^2} \Bigg\}_{+}\, \, , \quad j = 1, \ldots, p. $$
Since $\mu_j = 1/|\hat{\beta_j}|^2, \, j = 1, \ldots, p$, we have $\mu_j/ \big(|\hat{\beta_j}| \cdot \|X_j \|^2 \big) \asymp \sqrt{n}/(\log p)^{3/2} \gg 1$ if $j \in S_0^C$ which implies strong penalty for the noise component. Furthermore, if we assume that $|\beta_{0j}| > M$ for $j \in S$, then $\mu_j/ \big(|\hat{\beta_j}| \cdot \|X_j \|^2 \big) < 1$, implying mild penalty for the signals.

\section{Simulation Study} 
We consider a detailed simulation study to compare the operating characteristics of SAVS with various competitors. We considered model (\ref{model}) with $\sigma = 1.5$, $n \in \{ 100, 200 \}$ and $p \in \{500, 1000, 5000\}$. Rows of X were independently generated from $\mc N_p(0, \Sigma)$ with
\begin{itemize}
	\item[(i)] $\Sigma = I_p$ : Independent design
	\item[(ii)] $\Sigma_{jj} = 1, \quad \Sigma_{jj'} = 0.5, \quad j \neq j' = 1,2, \ldots, p$ : Compound symmetry
	\item[(iii)] $\Sigma_{jj'} = \rho^{|j - j'|}, \quad j, j' = 1,2, \ldots, p, \quad \text{with} \,\, \rho \in \{0.5, 0.7, 0.9\}$ : Toeplitz structure or AR(1).
\end{itemize}
For the number non-zero entries $s_0$ of the true regression coefficient $\beta_0$, we considered two choices viz. 5 and 10. 
\begin{itemize}
	\item[Case-1 :] The true $\beta_0$ had $s_0 = 5$ non-zero entries, with the non-zero entries having magnitude
	\begin{itemize}
		\item[] set-1 : $\{1.50, 1.75, 2.00, 2.25, 2.50\}$
		\item[] set-2 : $\{0.75, 1.00, 1.25, 1.50, 1.75\}$
	\end{itemize}
	multiplied by a random sign. 
	
	\item[Case-2 :] The true $\beta_0$ had $10$ non-zero entries corresponding to different simulation cases, with the non-zero entries $$\{0.75, 1.00, 1.25, 1.50, 1.75, 2.00, 2.25, 2.50, 2.75, 3.00\}$$ multiplied by a random sign. \\
\end{itemize}

We have 90 simulation cases corresponding to 90 parameter combinations altogether. For each case, we considered 1000 simulation replicates. 

We compared SAVS with smoothly clipped absolute deviation or SCAD \citep{fan2001variable}, minimax concave penalty or MCP \citep{zhang2010nearly} and Adaptive LASSO \citep{zou2006adaptive}. These three methods were implemented with help of \textbf{{\fontfamily{qcr} \selectfont R}} packages \textbf{{\fontfamily{qcr} \selectfont ncvreg}} and \textbf{{\fontfamily{qcr} \selectfont parcor}} based on 10-fold cross-validation. 
We additionally considered the recent maximum a posteriori (MAP) estimate named S5 \citep{shin2015scalable} implemented in the \textbf{{\fontfamily{qcr} \selectfont R}} package \textbf{{\fontfamily{qcr} \selectfont BayesS5}}, with default set-up. 

As a measure of performance, we used Matthew's Correlation Coefficient (MCC) defined as:
$$\text{MCC} = \frac{\text{TP} \times \text{TN} - \text{FP} \times \text{FN}}{\sqrt{(\text{TP} + \text{FP}) (\text{TP} + \text{FN}) (\text{TN} + \text{FP}) (\text{TN} + \text{FN})}}$$
where \text{TP}, \text{TN}, \text{FP} and \text{FN} correspond to True Positive, True Negative, False Positive and False Negative respectively.  MCC values lies between $-1 \,\text{and} \,+1$;
$\text{MCC} = 1$ corresponds to perfect classification, which in our case is equivalent to exactly estimating true signals and true noises. 
Thus, the estimation procedure which leads to $\text{MCC}$ values closer to 1 for most of the replicates and most of the simulations does a better job in variable selection.

We also considered the True Positive Rate (RPT) or Sensitivity, defined by $\text{TP}/(\text{TP} + \text{FN})$, True Negative Rate (TNR) or Specificity, defined by $\text{TN}/(\text{TN} + \text{FP})$, and proportion of times the exact true model was selected\footnote{As a default, both SCAD and MCP in \textbf{{\fontfamily{qcr} \selectfont R}} always selects an intercept; for a fair comparison, we didn't consider it as a false positive.}. 

In the following section, we provide detailed results for the case $s_0 = 10$.  The results corresponding to $s_0 = 5$ follow a similar pattern overall and are provided in the Appendix.

\subsection{\textbf{Results corresponding to $s_0 = 10$}}
\begin{figure}[H]
	\centering
	\includegraphics[width=\textwidth]{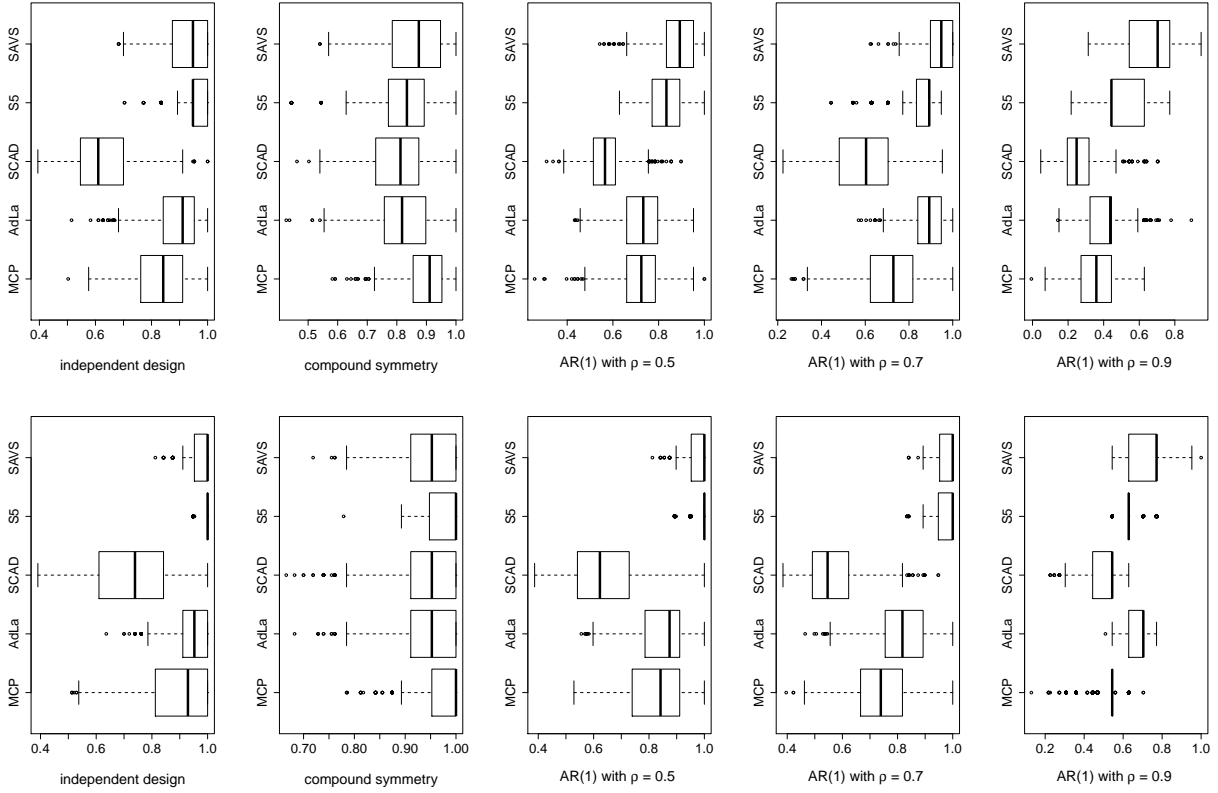}
	\caption{Boxplots of MCC values over 1000 replications for $s_0 = 10$ and $p = 500$ for the five methods; top and bottom rows correspond to $n = 100$ and $n = 200$; ``AdLa" is short for Adaptive LASSO.} \label{fig-p500-s10}
\end{figure}

\begin{figure}[H]
	\centering
	\includegraphics[width=\textwidth]{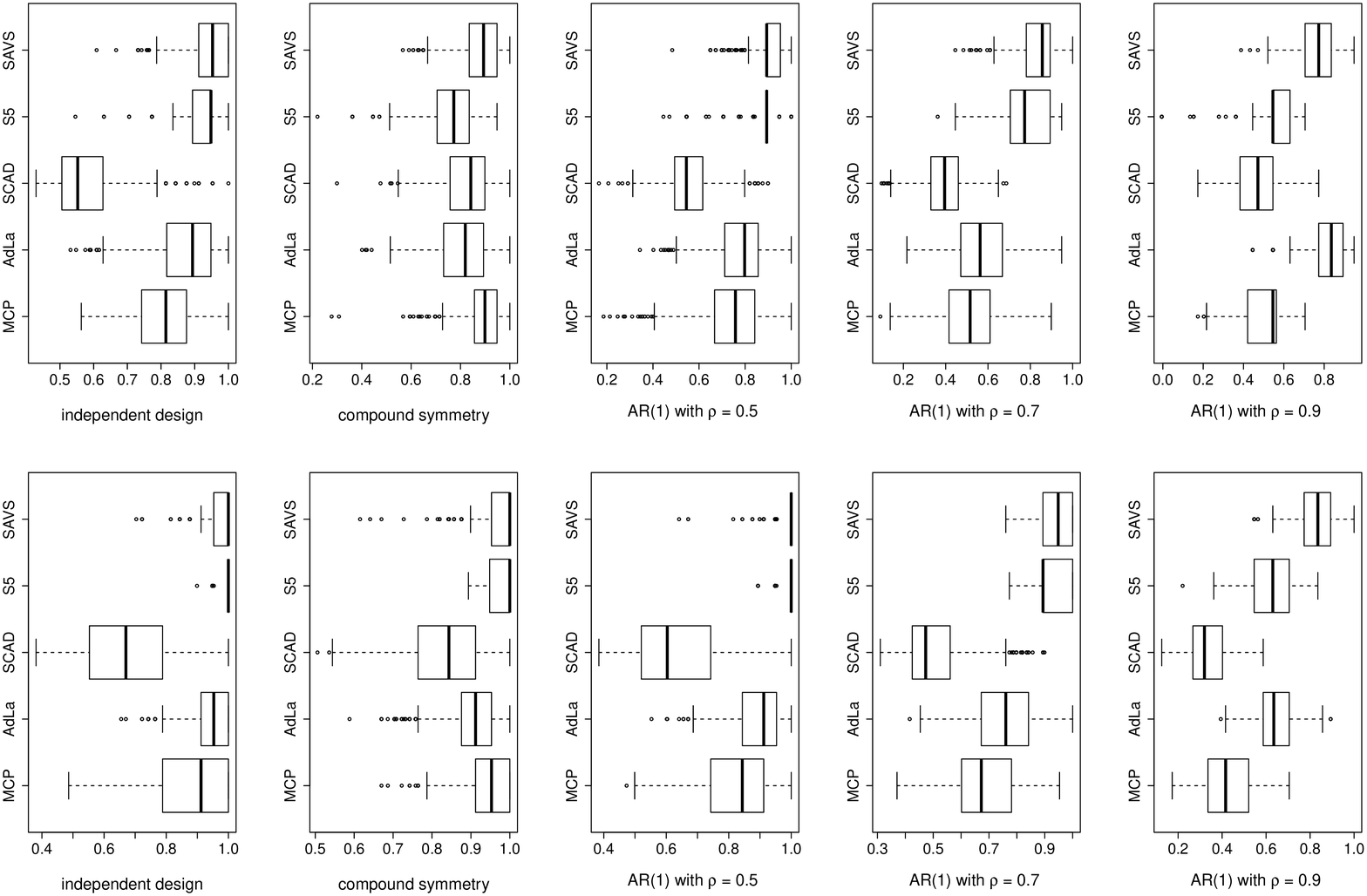}
	\caption{Boxplots of MCC values over 1000 replications for $s_0 = 10$ and $p = 1000$ for the five methods; top and bottom rows correspond to $n = 100$ and $n = 200$; ``AdLa" is short for Adaptive LASSO.} \label{fig-p1000-s10}
\end{figure}

\begin{figure}[H]
	\centering
	\includegraphics[width=\textwidth]{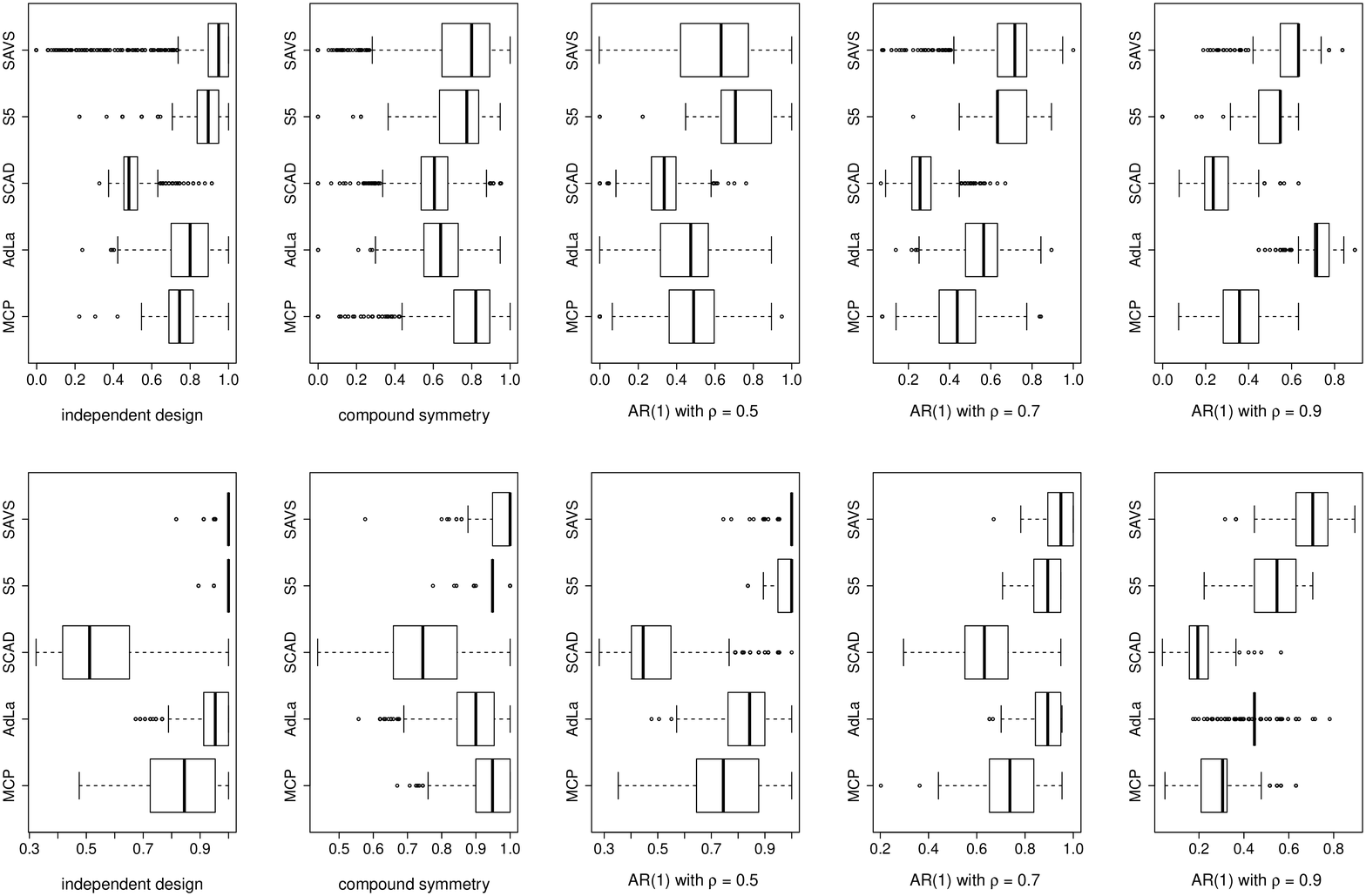}
	\caption{Boxplots of MCC values over 1000 replications for $s_0 = 10$ and $p = 5000$ for the five methods; top and bottom rows correspond to $n = 100$ and $n = 200$; ``AdLa" is short for Adaptive LASSO.} \label{fig-p5000-s10}
\end{figure}

The figures \ref{fig-p500-s10}, \ref{fig-p1000-s10} and \ref{fig-p5000-s10} report boxplots (displaying the five number summary: minimum, first quartile, median, third quartile, and maximum) of MCC values and the tables \ref{p500-s10}, \ref{p1000-s10} and \ref{p5000-s10} report an overall finding across the 1000 replicates for the given signal strengths, five designs, three values of $p$ and two values of $n$.

\begin{table}[H]
	\caption{Prop is the proportion of times true model being selected; for SCAD and MCP it corresponds to true model+intercept. Means and standard deviations(in subscript) for MCC, TPR and TNR over different methods are tabulated corresponding to $s_0 = 10$ and $p = 500$. ``AdLa" is short for Adaptive LASSO.}\label{p500-s10}
	\centering
	{\resizebox{\textwidth}{!}{\begin{tabular}{|c c||c c c c|c c c c|c c c c|c c c c|c c c c|}
				\hline
				\multirow{2}{*}{\parbox{0.07 \linewidth}{\centering $p$=500}} &  & \multicolumn{4}{|c|}{Independent} &  \multicolumn{4}{|c|}{Compound symmetry} & \multicolumn{4}{|c|}{\centering AR(1) with $\rho = 0.5$} & \multicolumn{4}{|c|}{\centering AR(1) with $\rho = 0.7$} & \multicolumn{4}{|c|}{\centering AR(1) with $\rho = 0.9$} \\ 
				& & Prop & MCC & TPR & TNR & Prop & MCC & TPR & TNR & Prop & MCC & TPR & TNR & Prop & MCC & TPR & TNR & Prop & MCC & TPR & TNR \\ 
				\hline 
				\multirow{4}{*}{\parbox{0.07 \linewidth}{\centering $n$=100}} & SAVS & 0.30 & $0.92_{\scriptstyle 0.07}$ & $0.99_{\scriptstyle 0.03}$ & $1.00_{\scriptstyle 0.00}$ & 0.09 & $0.85_{\scriptstyle 0.1}$ & $0.94_{\scriptstyle 0.07}$ & $1.00_{\scriptstyle 0.01}$ & 0.18 & $0.88_{\scriptstyle 0.09}$ & $0.89_{\scriptstyle 0.11}$ & $1.00_{\scriptstyle 0.00}$ & 0.31 & $0.94_{\scriptstyle 0.06}$ & $0.94_{\scriptstyle 0.06}$ & $1.00_{\scriptstyle 0.00}$ & 0.00 & $0.65_{\scriptstyle 0.14}$ & $0.45_{\scriptstyle 0.17}$ & $1.00_{\scriptstyle 0.00}$ \\ 
				& S5 & 0.28 & $0.95_{\scriptstyle 0.05}$ & $0.90_{\scriptstyle 0.08}$ & $1.00_{\scriptstyle 0.00}$ & 0.02 & $0.84_{\scriptstyle 0.09}$ & $0.72_{\scriptstyle 0.14}$ & $1.00_{\scriptstyle 0.00}$ & 0.08 & $0.85_{\scriptstyle 0.07}$ & $0.73_{\scriptstyle 0.12}$ & $0.92_{\scriptstyle 0.07}$ & 0.00 & $0.84_{\scriptstyle 0.09}$ & $0.71_{\scriptstyle 0.14}$ & $1.00_{\scriptstyle 0.00}$ & 0.00 & $0.51_{\scriptstyle 0.09}$ & $0.27_{\scriptstyle 0.10}$ & $1.00_{\scriptstyle 0.00}$ \\
				& SCAD & 0.00 & $0.63_{\scriptstyle 0.10}$ & $1.00_{\scriptstyle 0.02}$ & $0.97_{\scriptstyle 0.01}$ & 0.03 & $0.80_{\scriptstyle 0.09}$ & $0.96_{\scriptstyle 0.07}$ & $0.99_{\scriptstyle 0.01}$ & 0.00 & $0.57_{\scriptstyle 0.08}$ & $0.89_{\scriptstyle 0.11}$ & $0.97_{\scriptstyle 0.01}$ & 0.00 & $0.59_{\scriptstyle 0.15}$ & $0.77_{\scriptstyle 0.14}$ & $0.98_{\scriptstyle 0.01}$ & 0.00 & $0.27_{\scriptstyle 0.11}$ & $0.29_{\scriptstyle 0.11}$ & $0.98_{\scriptstyle 0.01}$ \\ 
				& AdLa & 0.18 & $0.89_{\scriptstyle 0.09}$ & $0.97_{\scriptstyle 0.05}$ & $0.99_{\scriptstyle 0.01}$ & 0.03 & $0.82_{\scriptstyle 0.1}$ & $0.9_{\scriptstyle 0.08}$ & $0.99_{\scriptstyle 0.01}$ & 0.00 & $0.73_{\scriptstyle 0.1}$ & $0.76_{\scriptstyle 0.1}$ & $0.99_{\scriptstyle 0.01}$ & 0.11 & $0.88_{\scriptstyle 0.08}$ & $0.9_{\scriptstyle 0.1}$ & $1.00_{\scriptstyle 0.00}$ & 0.00 & $0.40_{\scriptstyle 0.10}$ & $0.27_{\scriptstyle 0.11}$ & $0.99_{\scriptstyle 0.01}$ \\ 
				& MCP & 0.09 & $0.84_{\scriptstyle 0.10}$ & $0.99_{\scriptstyle 0.02}$ & $0.99_{\scriptstyle 0.01}$ & 0.13 & $0.90_{\scriptstyle 0.07}$ & $0.92_{\scriptstyle 0.09}$ & $1.00_{\scriptstyle 0.00}$ & 0.00 & $0.72_{\scriptstyle 0.10}$ & $0.85_{\scriptstyle 0.13}$ & $0.99_{\scriptstyle 0.01}$ & 0.00 & $0.70_{\scriptstyle 0.15}$ & $0.75_{\scriptstyle 0.14}$ & $0.99_{\scriptstyle 0.01}$ & 0.00 & $0.35_{\scriptstyle 0.09}$ & $0.25_{\scriptstyle 0.08}$ & $0.99_{\scriptstyle 0.01}$ \\ 
				\hline
				\multirow{4}{*}{\parbox{0.07 \linewidth}{\centering $n$=200}} & SAVS & 0.64 & $0.98_{\scriptstyle 0.03}$ & $1.00_{\scriptstyle 0.00}$ & $1.00_{\scriptstyle 0.00}$ & 0.4 & $0.95_{\scriptstyle 0.05}$ & $0.99_{\scriptstyle 0.02}$ & $0.99_{\scriptstyle 0.00}$ & 0.71 & $0.98_{\scriptstyle 0.03}$ & $1.00_{\scriptstyle 0.01}$ & $1.00_{\scriptstyle 0.00}$ & 0.72 & $0.98_{\scriptstyle 0.03}$ & $0.99_{\scriptstyle 0.04}$ & $1.00_{\scriptstyle 0.00}$ & 0.00 & $0.73_{\scriptstyle 0.08}$ & $0.55_{\scriptstyle 0.12}$ & $1.00_{\scriptstyle 0.00}$ \\ 
				& S5 & 0.96 & $1.00_{\scriptstyle 0.01}$ & $1.00_{\scriptstyle 0.02}$ & $1.00_{\scriptstyle 0.00}$ & 0.65 & $0.98_{\scriptstyle 0.03}$ & $0.96_{\scriptstyle 0.06}$ & $1.00_{\scriptstyle 0.00}$ & 0.92 & $0.99_{\scriptstyle 0.02}$ & $0.99_{\scriptstyle 0.04}$ & $1.00_{\scriptstyle 0.00}$ & 0.64 & $0.98_{\scriptstyle 0.04}$ & $0.96_{\scriptstyle 0.07}$ & $1.00_{\scriptstyle 0.00}$ & 0.00 & $0.63_{\scriptstyle 0.05}$ & $0.40_{\scriptstyle 0.07}$ & $1.00_{\scriptstyle 0.00}$ \\
				& SCAD & 0.04 & $0.73_{\scriptstyle 0.14}$ & $1.00_{\scriptstyle 0.00}$ & $0.98_{\scriptstyle 0.02}$ & 0.34 & $0.94_{\scriptstyle 0.07}$ & $1.00_{\scriptstyle 0.02}$ & $1.00_{\scriptstyle 0.00}$ & 0.00 & $0.64_{\scriptstyle 0.13}$ & $1.00_{\scriptstyle 0.03}$ & $0.97_{\scriptstyle 0.02}$ & 0.00 & $0.57_{\scriptstyle 0.10}$ & $0.97_{\scriptstyle 0.06}$ & $0.96_{\scriptstyle 0.02}$ & 0.00 & $0.49_{\scriptstyle 0.08}$ & $0.28_{\scriptstyle 0.05}$ & $1.00_{\scriptstyle 0.00}$ \\ 
				& AdLa & 0.44 & $0.94_{\scriptstyle 0.07}$ & $1.00_{\scriptstyle 0.01}$ & $0.99_{\scriptstyle 0.00}$ & 0.26 & $0.93_{\scriptstyle 0.06}$ & $0.99_{\scriptstyle 0.03}$ & $0.99_{\scriptstyle 0.00}$ & 0.09 & $0.85_{\scriptstyle 0.1}$ & $0.99_{\scriptstyle 0.03}$ & $0.99_{\scriptstyle 0.01}$ & 0.03 & $0.82_{\scriptstyle 0.1}$ & $0.95_{\scriptstyle 0.07}$ & $0.99_{\scriptstyle 0.01}$ & 0.00 & $0.68_{\scriptstyle 0.04}$ & $0.47_{\scriptstyle 0.05}$ & $1.00_{\scriptstyle 0.00}$ \\ 
				& MCP & 0.34 & $0.89_{\scriptstyle 0.12}$ & $1.00_{\scriptstyle 0.00}$ & $0.99_{\scriptstyle 0.01}$ & 0.58 & $0.97_{\scriptstyle 0.04}$ & $0.99_{\scriptstyle 0.03}$ & $1.00_{\scriptstyle 0.00}$ & 0.12 & $0.83_{\scriptstyle 0.12}$ & $1.00_{\scriptstyle 0.02}$ & $0.99_{\scriptstyle 0.01}$ & 0.01 & $0.74_{\scriptstyle 0.11}$ & $0.96_{\scriptstyle 0.07}$ & $0.98_{\scriptstyle 0.01}$ & 0.00 & $0.51_{\scriptstyle 0.07}$ & $0.29_{\scriptstyle 0.05}$ & $1.00_{\scriptstyle 0.00}$ \\ 
				\hline 
	\end{tabular}}}
\end{table}

\begin{table}[H]
	\caption{Prop is the proportion of times true model being selected; for SCAD and MCP it corresponds to true model+intercept. Means and standard deviations(in subscript) for MCC, TPR and TNR over different methods are tabulated corresponding to $s_0 = 10$ and $p = 1000$. ``AdLa" is short for Adaptive LASSO.}\label{p1000-s10}
	\centering
	{\resizebox{\textwidth}{!}{\begin{tabular}{|c c||c c c c|c c c c|c c c c|c c c c|c c c c|}\hline
				\multirow{2}{*}{\parbox{0.07 \linewidth}{\centering $p$=1000}} &  & \multicolumn{4}{|c|}{Independent} &  \multicolumn{4}{|c|}{Compound symmetry} & \multicolumn{4}{|c|}{\centering AR(1) with $\rho = 0.5$} & \multicolumn{4}{|c|}{\centering AR(1) with $\rho = 0.7$} & \multicolumn{4}{|c|}{\centering AR(1) with $\rho = 0.9$} \\ 
				& & Prop & MCC & TPR & TNR & Prop & MCC & TPR & TNR & Prop & MCC & TPR & TNR & Prop & MCC & TPR & TNR & Prop & MCC & TPR & TNR \\ 
				\hline 
				\multirow{4}{*}{\parbox{0.07 \linewidth}{\centering $n$=100}} & SAVS & 0.34 & $0.94_{\scriptstyle 0.06}$ & $0.98_{\scriptstyle 0.04}$ & $1.00_{\scriptstyle 0.00}$ & 0.08 & $0.88_{\scriptstyle 0.08}$ & $0.90_{\scriptstyle 0.08}$ & $1.00_{\scriptstyle 0.00}$ & 0.18 & $0.91_{\scriptstyle 0.07}$ & $0.89_{\scriptstyle 0.09}$ & $1.00_{\scriptstyle 0.00}$ & 0.02 & $0.84_{\scriptstyle 0.08}$ & $0.76_{\scriptstyle 0.12}$ & $1.00_{\scriptstyle 0.00}$ & 0.00 & $0.76_{\scriptstyle 0.08}$ & $0.59_{\scriptstyle 0.11}$ & $1.00_{\scriptstyle 0.00}$ \\ 
				& S5 & 0.16 & $0.93_{\scriptstyle 0.06}$ & $0.86_{\scriptstyle 0.10}$ & $1.00_{\scriptstyle 0.00}$ & 0.00 & $0.75_{\scriptstyle 0.12}$ & $0.58_{\scriptstyle 0.17}$ & $1.00_{\scriptstyle 0.00}$ & 0.02 & $0.87_{\scriptstyle 0.07}$ & $0.77_{\scriptstyle 0.10}$ & $1.00_{\scriptstyle 0.00}$ & 0.00 & $0.76_{\scriptstyle 0.12}$ & $0.60_{\scriptstyle 0.17}$ & $1.00_{\scriptstyle 0.00}$ & 0.00 & $0.58_{\scriptstyle 0.08}$ & $0.35_{\scriptstyle 0.07}$ & $1.00_{\scriptstyle 0.00}$ \\
				& SCAD & 0.00 & $0.58_{\scriptstyle 0.09}$ & $1.00_{\scriptstyle 0.02}$ & $0.98_{\scriptstyle 0.01}$ & 0.04 & $0.82_{\scriptstyle 0.10}$ & $0.91_{\scriptstyle 0.07}$ & $1.00_{\scriptstyle 0.00}$ & 0.00 & $0.56_{\scriptstyle 0.10}$ & $0.89_{\scriptstyle 0.09}$ & $0.98_{\scriptstyle 0.01}$ & 0.00 & $0.39_{\scriptstyle 0.1}$ & $0.64_{\scriptstyle 0.16}$ & $0.98_{\scriptstyle 0.01}$ & 0.00 & $0.46_{\scriptstyle 0.11}$ & $0.34_{\scriptstyle 0.08}$ & $1.00_{\scriptstyle 0.00}$ \\ 
				& AdLa & 0.12 & $0.87_{\scriptstyle 0.09}$ & $0.94_{\scriptstyle 0.07}$ & $0.99_{\scriptstyle 0.00}$ & 0.01 & $0.80_{\scriptstyle 0.10}$ & $0.85_{\scriptstyle 0.11}$ & $0.99_{\scriptstyle 0.00}$ & 0.01 & $0.78_{\scriptstyle 0.11}$ & $0.82_{\scriptstyle 0.1}$ & $0.99_{\scriptstyle 0.00}$ & 0.00 & $0.57_{\scriptstyle 0.13}$ & $0.5_{\scriptstyle 0.18}$ & $0.99_{\scriptstyle 0.00}$ & 0.00 & $0.82_{\scriptstyle 0.09}$ & $0.68_{\scriptstyle 0.14}$ & $1.00_{\scriptstyle 0.00}$ \\ 
				& MCP & 0.06 & $0.81_{\scriptstyle 0.10}$ & $0.99_{\scriptstyle 0.03}$ & $0.99_{\scriptstyle 0.00}$ & 0.12 & $0.90_{\scriptstyle 0.08}$ & $0.89_{\scriptstyle 0.09}$ & $1.00_{\scriptstyle 0.00}$ & 0.00 & $0.74_{\scriptstyle 0.12}$ & $0.84_{\scriptstyle 0.12}$ & $0.99_{\scriptstyle 0.00}$ & 0.00 & $0.51_{\scriptstyle 0.15}$ & $0.54_{\scriptstyle 0.16}$ & $0.99_{\scriptstyle 0.00}$ & 0.00 & $0.50_{\scriptstyle 0.10}$ & $0.33_{\scriptstyle 0.06}$ & $1.00_{\scriptstyle 0.00}$ \\ 
				\hline 
				\multirow{4}{*}{\parbox{0.07 \linewidth}{\centering $n$=200}} & SAVS & 0.73 & $0.98_{\scriptstyle 0.03}$ & $1.00_{\scriptstyle 0.00}$ & $1.00_{\scriptstyle 0.00}$ & 0.54 & $0.98_{\scriptstyle 0.04}$ & $0.99_{\scriptstyle 0.03}$ & $1.00_{\scriptstyle 0.00}$ & 0.76 & $0.99_{\scriptstyle 0.03}$ & $1.00_{\scriptstyle 0.01}$ & $1.00_{\scriptstyle 0.00}$ & 0.39 & $0.94_{\scriptstyle 0.06}$ & $0.90_{\scriptstyle 0.10}$ & $1.00_{\scriptstyle 0.00}$ & 0.00 & $0.83_{\scriptstyle 0.07}$ & $0.70_{\scriptstyle 0.11}$ & $1.00_{\scriptstyle 0.00}$ \\ 
				& S5 & 0.93 & $1.00_{\scriptstyle 0.01}$ & $0.99_{\scriptstyle 0.03}$ & $1.00_{\scriptstyle 0.00}$ & 0.55 & $0.97_{\scriptstyle 0.04}$ & $0.95_{\scriptstyle 0.07}$ & $1.00_{\scriptstyle 0.00}$ & 0.78 & $0.99_{\scriptstyle 0.02}$ & $0.98_{\scriptstyle 0.04}$ & $1.00_{\scriptstyle 0.00}$ & 0.33 & $0.93_{\scriptstyle 0.06}$ & $0.86_{\scriptstyle 0.10}$ & $1.00_{\scriptstyle 0.00}$ & 0.00 & $0.62_{\scriptstyle 0.08}$ & $0.39_{\scriptstyle 0.10}$ & $1.00_{\scriptstyle 0.00}$ \\
				& SCAD & 0.01 & $0.68_{\scriptstyle 0.15}$ & $1.00_{\scriptstyle 0.00}$ & $0.98_{\scriptstyle 0.01}$ & 0.07 & $0.83_{\scriptstyle 0.10}$ & $0.99_{\scriptstyle 0.02}$ & $0.99_{\scriptstyle 0.00}$ & 0.01 & $0.63_{\scriptstyle 0.14}$ & $1.00_{\scriptstyle 0.01}$ & $0.98_{\scriptstyle 0.01}$ & 0.00 & $0.50_{\scriptstyle 0.11}$ & $0.89_{\scriptstyle 0.09}$ & $0.97_{\scriptstyle 0.01}$ & 0.00 & $0.34_{\scriptstyle 0.10}$ & $0.45_{\scriptstyle 0.13}$ & $0.98_{\scriptstyle 0.01}$ \\ 
				& AdLa & 0.47 & $0.95_{\scriptstyle 0.06}$ & $1.00_{\scriptstyle 0.01}$ & $1.00_{\scriptstyle 0.00}$ & 0.18 & $0.91_{\scriptstyle 0.07}$ & $0.99_{\scriptstyle 0.03}$ & $0.99_{\scriptstyle 0.00}$ & 0.17 & $0.89_{\scriptstyle 0.08}$ & $0.99_{\scriptstyle 0.02}$ & $0.99_{\scriptstyle 0.00}$ & 0.00 & $0.75_{\scriptstyle 0.1}$ & $0.83_{\scriptstyle 0.06}$ & $0.99_{\scriptstyle 0.00}$ & 0.00 & $0.64_{\scriptstyle 0.09}$ & $0.61_{\scriptstyle 0.12}$ & $0.99_{\scriptstyle 0.00}$ \\ 
				& MCP & 0.28 & $0.87_{\scriptstyle 0.13}$ & $1.00_{\scriptstyle 0.00}$ & $1.00_{\scriptstyle 0.01}$ & 0.43 & $0.95_{\scriptstyle 0.06}$ & $0.99_{\scriptstyle 0.03}$ & $1.00_{\scriptstyle 0.00}$ & 0.12 & $0.83_{\scriptstyle 0.12}$ & $1.00_{\scriptstyle 0.02}$ & $0.99_{\scriptstyle 0.01}$ & 0.00 & $0.69_{\scriptstyle 0.12}$ & $0.85_{\scriptstyle 0.10}$ & $0.99_{\scriptstyle 0.01}$ & 0.00 & $0.42_{\scriptstyle 0.10}$ & $0.39_{\scriptstyle 0.11}$ & $0.99_{\scriptstyle 0.01}$ \\ 
				\hline 
	\end{tabular}}}
\end{table}

\begin{table}[H]
	\caption{Prop is the proportion of times true model being selected; for SCAD and MCP it corresponds to true model+intercept. Means and standard deviations(in subscript) for MCC, TPR and TNR over different methods are tabulated corresponding to $s_0 = 10$ and $p = 5000$. ``AdLa" is short for Adaptive LASSO.}\label{p5000-s10}
	\centering
	{\resizebox{\textwidth}{!}{\begin{tabular}{|c c||c c c c|c c c c|c c c c|c c c c|c c c c|}\hline
				\multirow{2}{*}{\parbox{0.07 \linewidth}{\centering $p$=5000}} &  & \multicolumn{4}{|c|}{Independent} &  \multicolumn{4}{|c|}{Compound symmetry} & \multicolumn{4}{|c|}{\centering AR(1) with $\rho = 0.5$} & \multicolumn{4}{|c|}{\centering AR(1) with $\rho = 0.7$} & \multicolumn{4}{|c|}{\centering AR(1) with $\rho = 0.9$} \\ 
				& & Prop & MCC & TPR & TNR & Prop & MCC & TPR & TNR & Prop & MCC & TPR & TNR & Prop & MCC & TPR & TNR & Prop & MCC & TPR & TNR \\ 
				\hline 
				\multirow{4}{*}{\parbox{0.07 \linewidth}{\centering $n$=100}} & SAVS & 0.26 & $0.87_{\scriptstyle 0.20}$ & $0.85_{\scriptstyle 0.20}$ & $1.00_{\scriptstyle 0.00}$ & 0.02 & $0.74_{\scriptstyle 0.21}$ & $0.67_{\scriptstyle 0.21}$ & $1.00_{\scriptstyle 0.00}$ & 0.00 & $0.58_{\scriptstyle 0.27}$ & $0.48_{\scriptstyle 0.23}$ & $1.00_{\scriptstyle 0.00}$ & 0.00 & $0.70_{\scriptstyle 0.15}$ & $0.56_{\scriptstyle 0.15}$ & $1.00_{\scriptstyle 0.00}$ & 0.00 & $0.59_{\scriptstyle 0.09}$ & $0.38_{\scriptstyle 0.08}$ & $1.00_{\scriptstyle 0.00}$ \\ 
				& S5 & 0.02 & $0.86_{\scriptstyle 0.10}$ & $0.75_{\scriptstyle 0.16}$ & $1.00_{\scriptstyle 0.00}$ & 0.02 & $0.70_{\scriptstyle 0.16}$ & $0.53_{\scriptstyle 0.21}$ & $1.00_{\scriptstyle 0.00}$ & 0.00 & $0.72_{\scriptstyle 0.13}$ & $0.54_{\scriptstyle 0.19}$ & $1.00_{\scriptstyle 0.00}$ & 0.00 & $0.68_{\scriptstyle 0.08}$ & $0.48_{\scriptstyle 0.11}$ & $1.00_{\scriptstyle 0.00}$ & 0.00 & $0.50_{\scriptstyle 0.07}$ & $0.26_{\scriptstyle 0.06}$ & $1.00_{\scriptstyle 0.00}$ \\
				& SCAD & 0.00 & $0.50_{\scriptstyle 0.07}$ & $0.99_{\scriptstyle 0.03}$ & $0.99_{\scriptstyle 0.00}$ & 0.00 & $0.61_{\scriptstyle 0.13}$ & $0.86_{\scriptstyle 0.16}$ & $1.00_{\scriptstyle 0.00}$ & 0.00 & $0.33_{\scriptstyle 0.11}$ & $0.61_{\scriptstyle 0.18}$ & $0.99_{\scriptstyle 0.00}$ & 0.00 & $0.27_{\scriptstyle 0.08}$ & $0.46_{\scriptstyle 0.09}$ & $0.99_{\scriptstyle 0.00}$ & 0.00 & $0.26_{\scriptstyle 0.09}$ & $0.28_{\scriptstyle 0.06}$ & $1.00_{\scriptstyle 0.00}$ \\ 
				& AdLa & 0.02 & $0.78_{\scriptstyle 0.13}$ & $0.8_{\scriptstyle 0.13}$ & $1.00_{\scriptstyle 0.00}$ & 0.00 & $0.63_{\scriptstyle 0.14}$ & $0.68_{\scriptstyle 0.16}$ & $1.00_{\scriptstyle 0.00}$ & 0.00 & $0.43_{\scriptstyle 0.20}$ & $0.39_{\scriptstyle 0.20}$ & $1.00_{\scriptstyle 0.00}$ & 0.00 & $0.58_{\scriptstyle 0.11}$ & $0.46_{\scriptstyle 0.12}$ & $1.00_{\scriptstyle 0.00}$ & 0.00 & $0.73_{\scriptstyle 0.08}$ & $0.57_{\scriptstyle 0.10}$ & $1.00_{\scriptstyle 0.00}$ \\ 
				& MCP & 0.01 & $0.76_{\scriptstyle 0.09}$ & $0.98_{\scriptstyle 0.05}$ & $1.00_{\scriptstyle 0.00}$ & 0.02 & $0.77_{\scriptstyle 0.18}$ & $0.75_{\scriptstyle 0.21}$ & $1.00_{\scriptstyle 0.00}$ & 0.00 & $0.47_{\scriptstyle 0.18}$ & $0.52_{\scriptstyle 0.20}$ & $1.00_{\scriptstyle 0.00}$ & 0.00 & $0.44_{\scriptstyle 0.12}$ & $0.41_{\scriptstyle 0.08}$ & $1.00_{\scriptstyle 0.00}$ & 0.00 & $0.36_{\scriptstyle 0.11}$ & $0.27_{\scriptstyle 0.05}$ & $1.00_{\scriptstyle 0.00}$ \\ 
				\hline 
				\multirow{4}{*}{\parbox{0.07 \linewidth}{\centering $n$=200}} & SAVS & 0.93 & $1.00_{\scriptstyle 0.02}$ & $1.00_{\scriptstyle 0.02}$ & $1.00_{\scriptstyle 0.00}$ & 0.56 & $0.97_{\scriptstyle 0.04}$ & $0.96_{\scriptstyle 0.05}$ & $1.00_{\scriptstyle 0.00}$ & 0.82 & $0.99_{\scriptstyle 0.02}$ & $0.98_{\scriptstyle 0.04}$ & $1.00_{\scriptstyle 0.00}$ & 0.34 & $0.94_{\scriptstyle 0.05}$ & $0.9_{\scriptstyle 0.09}$ & $1.00_{\scriptstyle 0.00}$ & 0.00 & $0.70_{\scriptstyle 0.08}$ & $0.51_{\scriptstyle 0.10}$ & $1.00_{\scriptstyle 0.00}$ \\ 
				& S5 & 0.78 & $0.99_{\scriptstyle 0.02}$ & $0.98_{\scriptstyle 0.07}$ & $1.00_{\scriptstyle 0.00}$ & 0.19 & $0.95_{\scriptstyle 0.03}$ & $0.91_{\scriptstyle 0.05}$ & $1.00_{\scriptstyle 0.00}$ & 0.52 & $0.97_{\scriptstyle 0.03}$ & $0.95_{\scriptstyle 0.06}$ & $1.00_{\scriptstyle 0.00}$ & 0.00 & $0.89_{\scriptstyle 0.05}$ & $0.79_{\scriptstyle 0.10}$ & $1.00_{\scriptstyle 0.00}$ & 0.00 & $0.52_{\scriptstyle 0.12}$ & $0.29_{\scriptstyle 0.09}$ & $1.00_{\scriptstyle 0.00}$ \\
				& SCAD & 0.00 & $0.54_{\scriptstyle 0.15}$ & $1.00_{\scriptstyle 0.00}$ & $0.99_{\scriptstyle 0.00}$ & 0.02 & $0.75_{\scriptstyle 0.12}$ & $0.99_{\scriptstyle 0.03}$ & $1.00_{\scriptstyle 0.00}$ & 0.00 & $0.49_{\scriptstyle 0.13}$ & $0.99_{\scriptstyle 0.02}$ & $0.99_{\scriptstyle 0.00}$ & 0.00 & $0.64_{\scriptstyle 0.12}$ & $0.80_{\scriptstyle 0.07}$ & $1.00_{\scriptstyle 0.00}$ & 0.00 & $0.20_{\scriptstyle 0.07}$ & $0.32_{\scriptstyle 0.13}$ & $0.99_{\scriptstyle 0.00}$ \\ 
				& AdLa & 0.46 & $0.95_{\scriptstyle 0.08}$ & $0.99_{\scriptstyle 0.02}$ & $1.00_{\scriptstyle 0.00}$ & 0.13 & $0.88_{\scriptstyle 0.08}$ & $0.96_{\scriptstyle 0.05}$ & $1.00_{\scriptstyle 0.00}$ & 0.06 & $0.83_{\scriptstyle 0.10}$ & $0.89_{\scriptstyle 0.10}$ & $1.00_{\scriptstyle 0.00}$ & 0.00 & $0.89_{\scriptstyle 0.05}$ & $0.84_{\scriptstyle 0.05}$ & $1.00_{\scriptstyle 0.00}$ & 0.00 & $0.45_{\scriptstyle 0.07}$ & $0.24_{\scriptstyle 0.07}$ & $1.00_{\scriptstyle 0.00}$ \\ 
				& MCP & 0.17 & $0.83_{\scriptstyle 0.14}$ & $1.00_{\scriptstyle 0.00}$ & $1.00_{\scriptstyle 0.00}$ & 0.29 & $0.93_{\scriptstyle 0.06}$ & $0.98_{\scriptstyle 0.04}$ & $1.00_{\scriptstyle 0.00}$ & 0.04 & $0.75_{\scriptstyle 0.14}$ & $0.99_{\scriptstyle 0.04}$ & $1.00_{\scriptstyle 0.00}$ & 0.00 & $0.73_{\scriptstyle 0.11}$ & $0.77_{\scriptstyle 0.07}$ & $1.00_{\scriptstyle 0.00}$ & 0.00 & $0.28_{\scriptstyle 0.10}$ & $0.26_{\scriptstyle 0.13}$ & $1.00_{\scriptstyle 0.00}$ \\ 
				\hline 
	\end{tabular}}}
\end{table}

\subsection{\textbf{Summary of results}} 

The results overall indicate that SAVS is highly competitive to the existing methods across all the settings.  In terms of the proportion of times of identifying the correct model, SAVS was best or second best 73 (out of 90) times. In terms of MCC, SAVS was best or second best 86 (out of 90) times. Similar conclusions can be drawn based on TNR and TPR values. 

In particular, SAVS had an overall superior performance in the correlated design setting. Among the frequentist competitors, adaptive Lasso was the closest to SAVS. S5 performed strongly under independent design and mild correlations, though its performance somewhat deteriorated with increasing correlation.

\section{Application on a real data set}

We considered the dataset studied by \cite{lan2006combined} on coordinated regulation of gene expression levels recorded for 31 female and 29 male mice (i.e, 60 subjects in total). The dataset is consists of $22,575$ gene expression values along with a number of psychological phenotypes, including numbers of stearoyl-CoA desaturase 1 (\textbf{SCD1}), glycerol-3-phosphate acyltransferase (\textbf{GPAT}) and phos-phoenopyruvate carboxykinase (\textbf{PEPCK}). The last three quantities are response variables and were measured by quantitative real-time RT-PCR. This data set is publicly available at \href{URL}{http://www.ncbi.nlm.nih.gov/geo} (accession number GSE3330). The main purpose of the analysis here was to show that SAVS can scale up to real problems involving a large number of predictors.

\begin{table}[H]
	\caption{The table shows the indices and probe set ID of the selected genes using SAVS and S5 for GPAT, PEPCK and SCD1}\label{real-data}
	\centering
	\begin{tabular}{|c ||c c|c c|}\hline
		\multirow{2}{*}{Phenotypes} & \multicolumn{2}{|c|}{SAVS} &  \multicolumn{2}{|c|}{S5} \\ 
		& Index & probe set ID & Index & probe set ID \\ 
		\hline
		\multirow{2}{*}{\parbox{0.1 \linewidth} {\centering GPAT}} & 10854 & 1442439\textunderscore at & 17498 & 1454280\textunderscore at \\
		&  &  & 18639 & 1455980\textunderscore a\textunderscore at \\
		\hline
		\multirow{2}{*}{\parbox{0.1 \linewidth} {\centering PEPCK}} & 7640 & 1438937\textunderscore x\textunderscore at & 7640 & 1438937\textunderscore x\textunderscore at \\
		& 18263 & 1455375\textunderscore at & 18558 & 1455816\textunderscore a\textunderscore at \\
		\hline
		\multirow{2}{*}{\parbox{0.1 \linewidth} {\centering SCD1}} & 6002 & 1436216\textunderscore s\textunderscore at & 4664 & 1434072\textunderscore at \\
		& 10310 & 1441881\textunderscore x\textunderscore at & 4729 & 1434185\textunderscore at \\
		\hline
	\end{tabular}
\end{table}

We applied SAVS and S5 on this data for all the three responses. Table \ref{real-data} gives the probe set ID \citep{lan2006combined} of the selected genes by each method.

\section{Discussion} 

The literature on Bayesian sparse shrinkage priors has continued to impressively grow over the past decade and half. In this article, we have proposed a simple method called SAVS to summarize the posterior distribution from such shrinkage priors to obtain a set of selected variables. While we focus on the horseshoe prior, the method extends trivially to other priors. The SAVS approach for the horsehsoe prior had impressive operating characteristics across a wide range of simulations. 

While we haven't explored it here, the proposed approach can additionally be used to obtain a characterization of uncertainty in variable selection by applying the SAVS procedure to each MCMC iterate rather than the posterior mean. We leave this topic for future investigation. Another potential attraction of the proposed approach is the almost immediate generalization to related high-dimensional models, such as glm regression, factor regression, and tensor regression, to name a few.

\section{R code}

R code to implement the simulations for $s_0 = 10$ are provided in \href{URL}{https://github.com/raypallavi/SAVS.git}

\section{APPENDIX}
\textbf{\Large Results corresponding to $s_0 = 5$ :}\\

The tables \ref{tab-p500-n200}, \ref{tab-p1000-n200}, \ref{tab-p5000-n200}, \ref{tab-p500-n100}, \ref{tab-p1000-n100} and \ref{tab-p5000-n100} report an overall finding across the 1000 replicates for the two signal strengths, five designs, three values of $p$ and two values of $n$.

\begin{figure}[H]
	\centering
	\includegraphics[width=\textwidth]{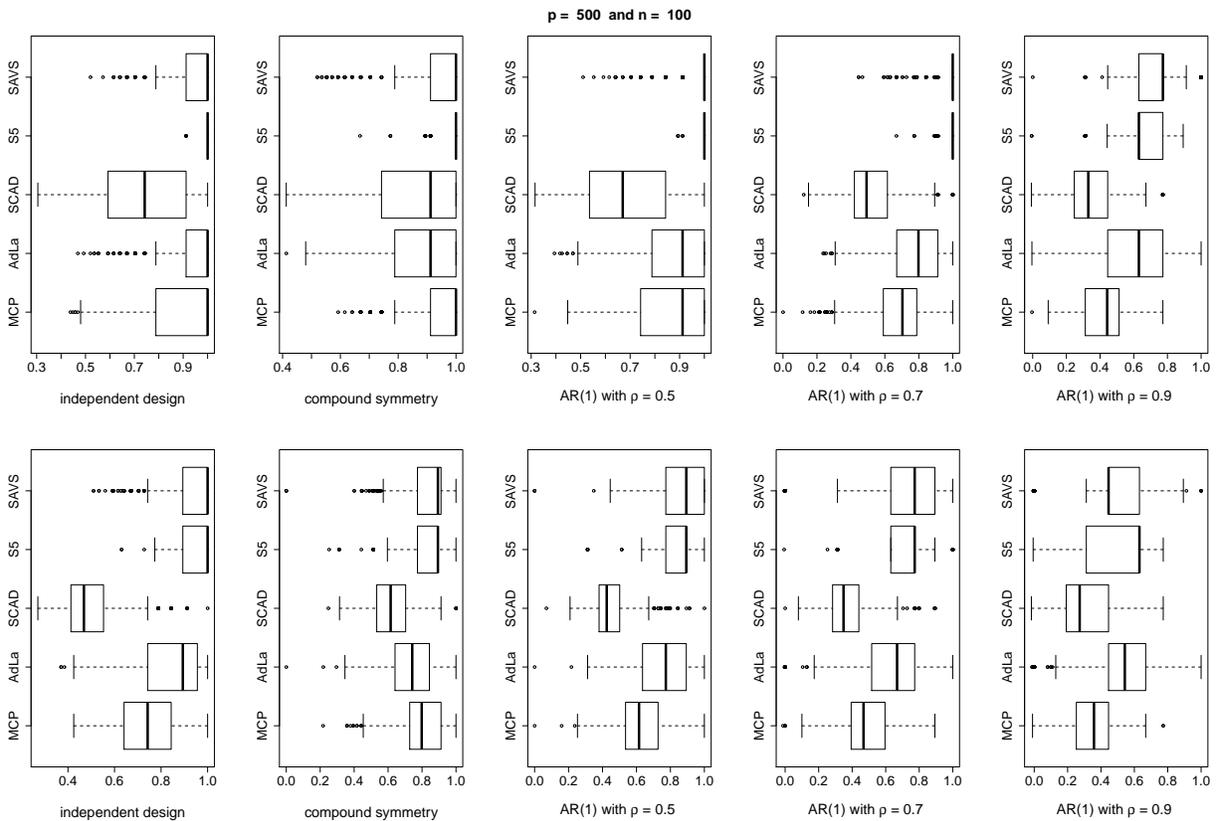}
	\caption{Boxplots of MCC values over 1000 replications for $s_0 = 5$ for the five methods; top and bottom rows correspond to the set-1 and set-2 values of $\beta_0$; ``AdLa" is short for Adaptive LASSO} \label{fig-p500-n100}
\end{figure}

\begin{figure}[H]
	\centering
	\includegraphics[width=\textwidth]{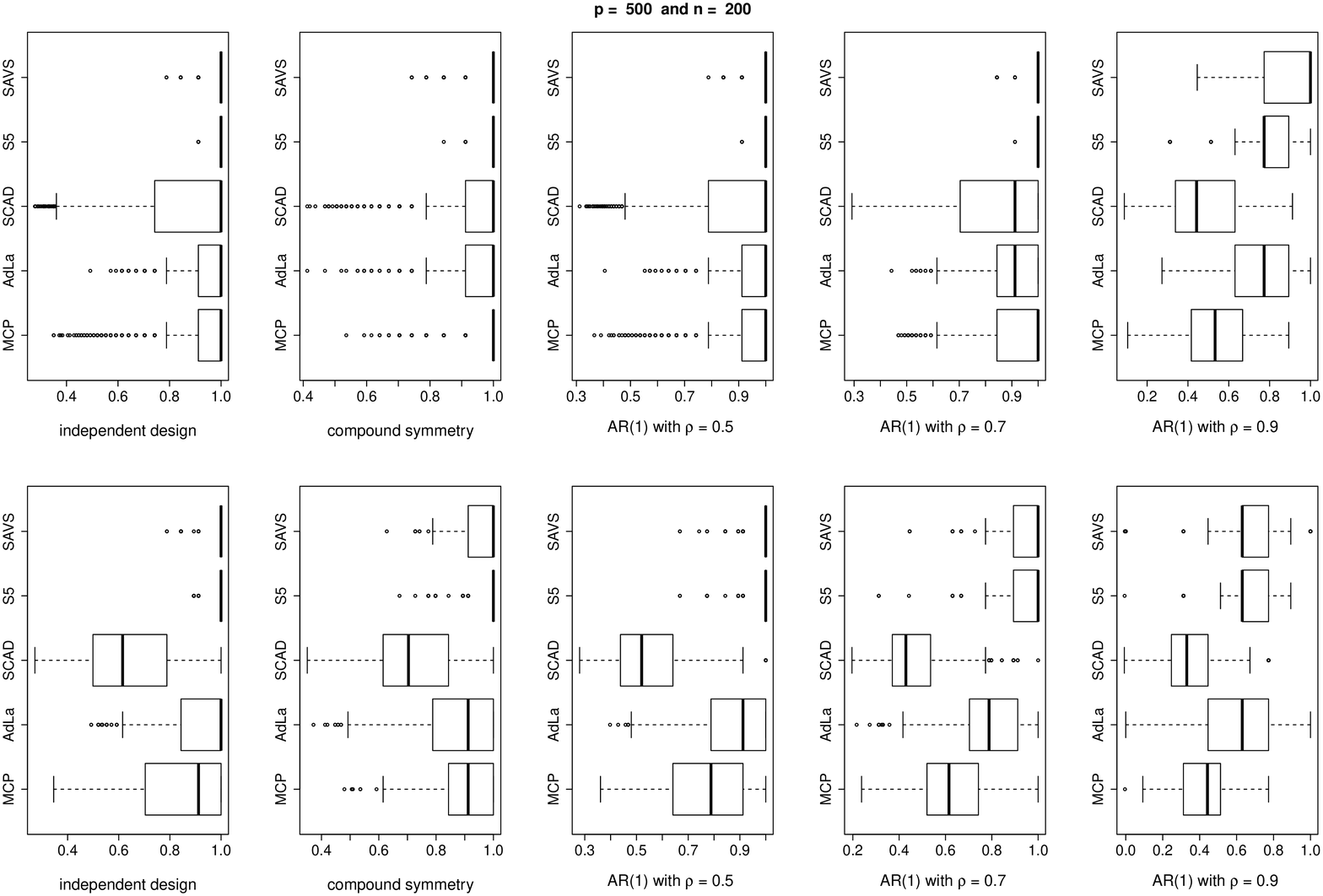}
	\caption{Boxplots of MCC values over 1000 replications for $s_0 = 5$ for the five methods; top and bottom rows correspond to the set-1 and set-2 values of $\beta_0$; ``AdLa" is short for Adaptive LASSO } \label{fig-p500-n200}
\end{figure}

\begin{figure}[H]
	\centering
	\includegraphics[width=\textwidth]{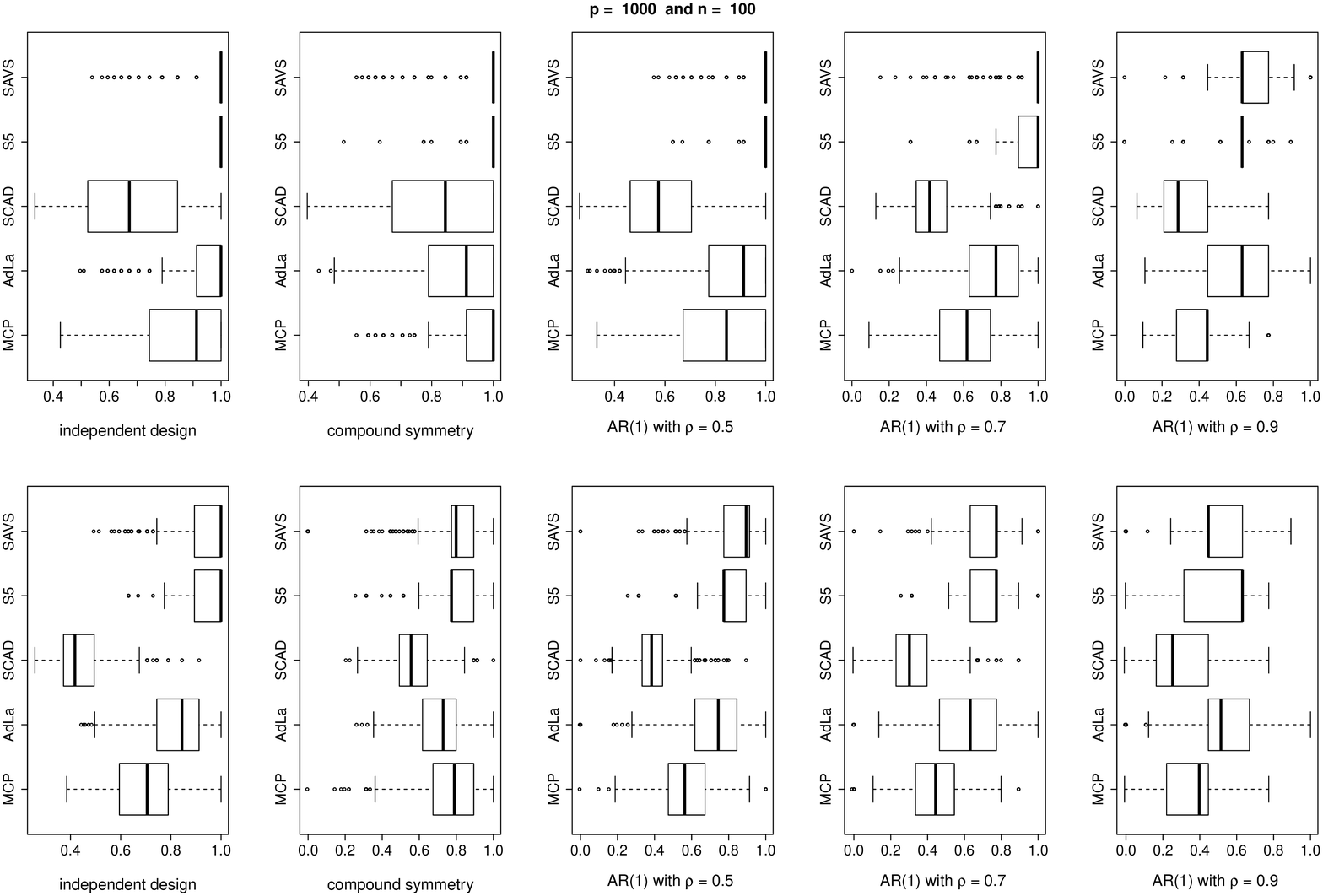}
	\caption{Boxplots of MCC values over 1000 replications for $s_0 = 5$ for the five methods; top and bottom rows correspond to the set-1 and set-2 values of $\beta_0$; ``AdLa" is short for Adaptive LASSO} \label{fig-p1000-n100}
\end{figure}

\begin{figure}[H]
	\centering
	\includegraphics[width=\textwidth]{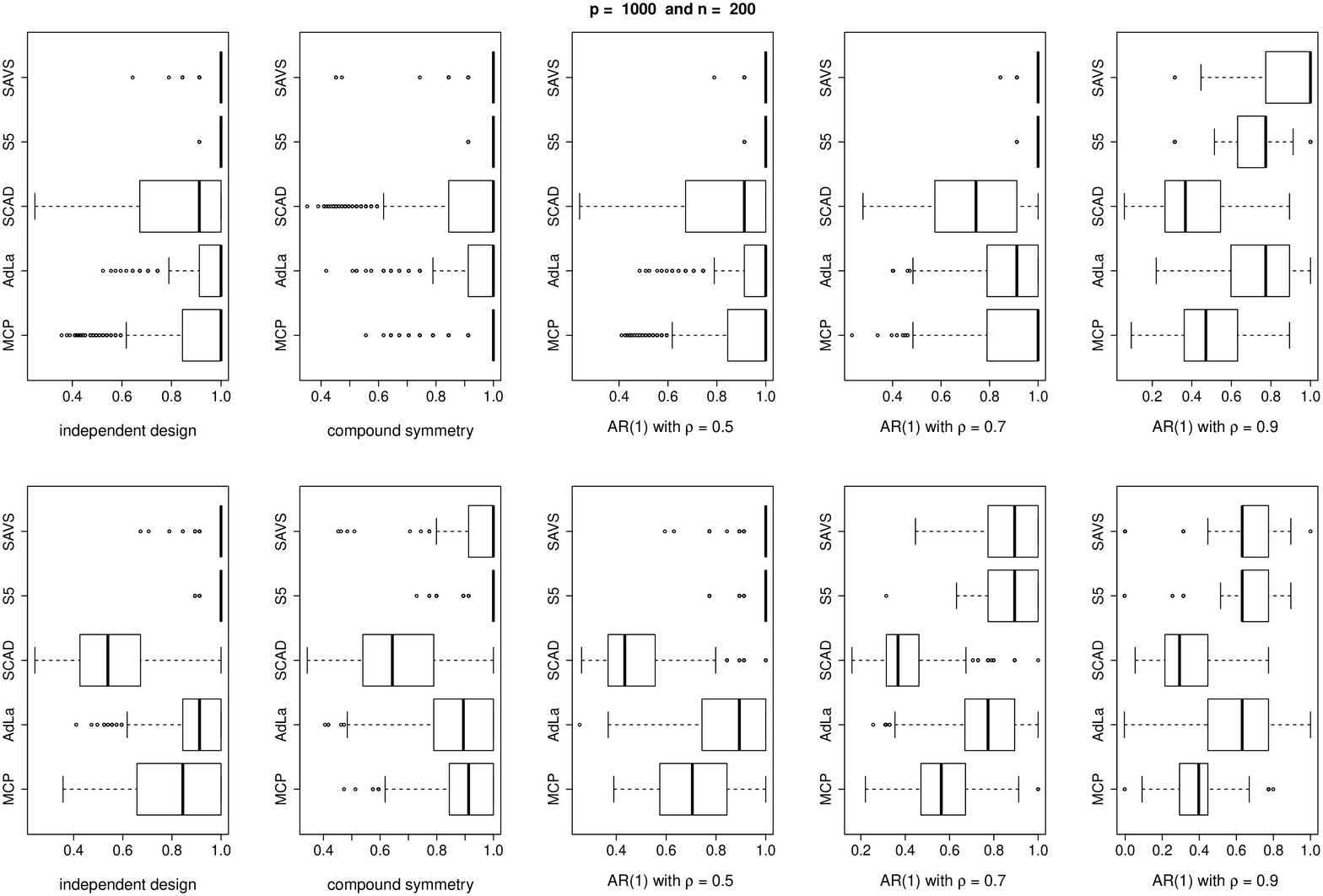}
	\caption{Boxplots of MCC values over 1000 replications for $s_0 = 5$ for the five methods; top and bottom rows correspond to the set-1 and set-2 values of $\beta_0$; ``AdLa" is short for Adaptive LASSO } \label{fig-p1000-n200}
\end{figure}

\begin{figure}[H]
	\centering
	\includegraphics[width=\textwidth]{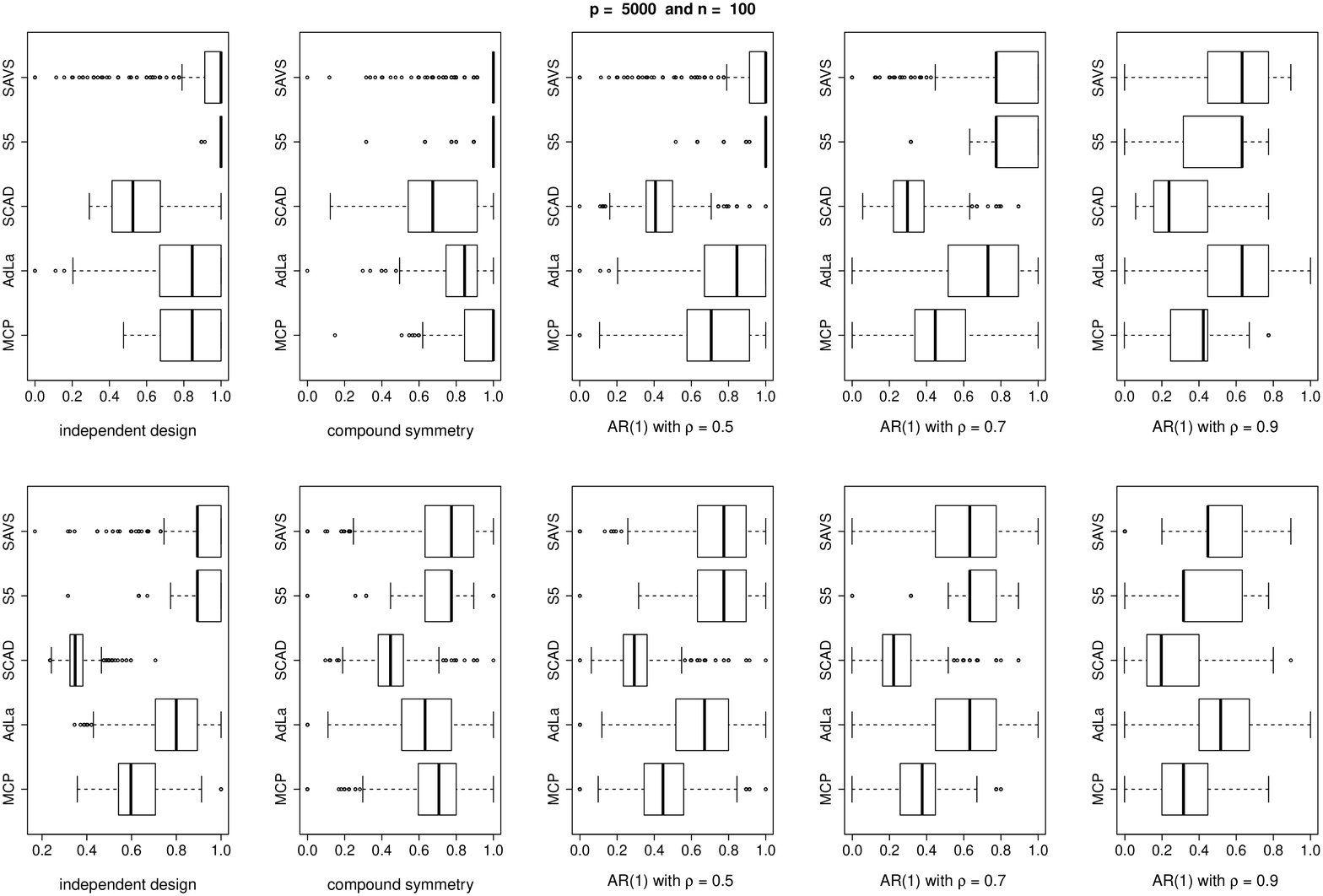}
	\caption{Boxplots of MCC values over 1000 replications for $s_0 = 5$ for the five methods; top and bottom rows correspond to the set-1 and set-2 values of $\beta_0$; ``AdLa" is short for Adaptive LASSO} \label{fig-p5000-n100}
\end{figure}

\begin{figure}[H]
	\centering
	\includegraphics[width=\textwidth]{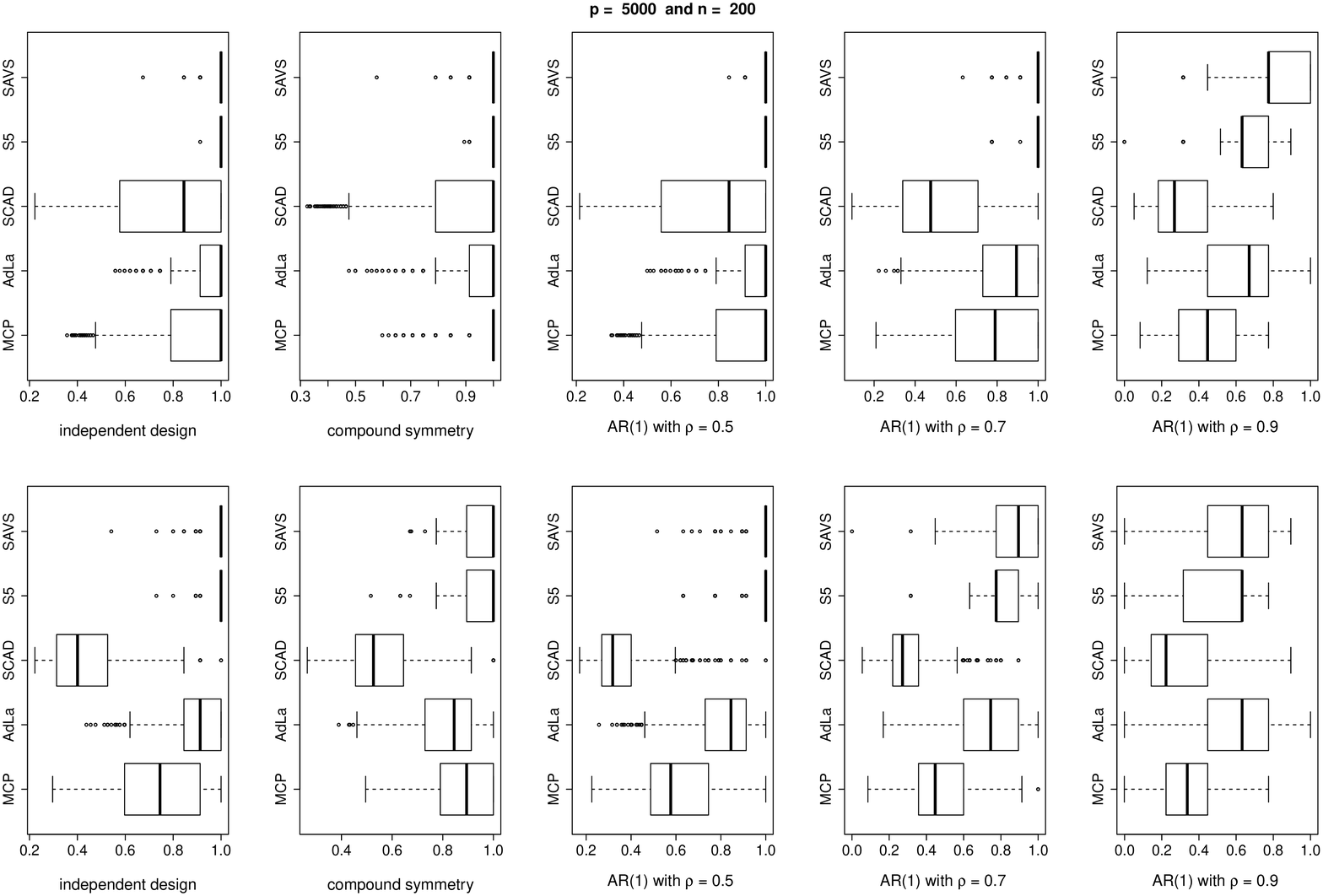}
	\caption{Boxplots of MCC values over 1000 replications for $s_0 = 5$ for the five methods; top and bottom rows correspond to the set-1 and set-2 values of $\beta_0$; ``AdLa" is short for Adaptive LASSO} \label{fig-p5000-n200}
\end{figure}

Figures \ref{fig-p500-n100}, \ref{fig-p500-n200}, \ref{fig-p1000-n100}, \ref{fig-p1000-n200}, \ref{fig-p5000-n100} and \ref{fig-p5000-n200} report boxplots (displaying the five number summary: minimum, first quartile, median, third quartile, and maximum) of MCC values corresponding to $s_0 = 5$.

\begin{table}[H]
	\caption{Prop is the proportion of times true model being selected; for SCAD and MCP it corresponds to true model+intercept. Means and standard deviations(in subscript) for MCC, TPR and TNR over different methods are tabulated corresponding to $s_0 = 5$, $p = 500$ and $n = 100$. ``AdLa" is short for Adaptive LASSO}\label{tab-p500-n100}
	\centering
	{\resizebox{\textwidth}{!}{\begin{tabular}{|c c||c c c c|c c c c|c c c c|c c c c|c c c c|}\hline
				\multirow{2}{*}{\parbox{0.07 \linewidth}{$p$=500 \ $n$=100}} &  & \multicolumn{4}{|c|}{Independent} &  \multicolumn{4}{|c|}{Compound symmetry} & \multicolumn{4}{|c|}{\centering AR(1) with $\rho = 0.5$} & \multicolumn{4}{|c|}{\centering AR(1) with $\rho = 0.7$} & \multicolumn{4}{|c|}{\centering AR(1) with $\rho = 0.9$} \\ 
				& & Prop & MCC & TPR & TNR & Prop & MCC & TPR & TNR & Prop & MCC & TPR & TNR & Prop & MCC & TPR & TNR & Prop & MCC & TPR & TNR \\ 
				\hline 
				\multirow{4}{*}{\parbox{0.07 \linewidth}{\centering Set-1}} & SAVS & 0.72 & $0.96_{\scriptstyle 0.08}$ & $1.00_{\scriptstyle 0.00}$ & $1.00_{\scriptstyle 0.00}$ & 0.64 & $0.95_{\scriptstyle 0.09}$ & $1.00_{\scriptstyle 0.01}$ & $1.00_{\scriptstyle 0.00}$ & 0.81 & $0.97_{\scriptstyle 0.06}$ & $1.00_{\scriptstyle 0.01}$ & $1.00_{\scriptstyle 0.00}$ & 0.81 & $0.97_{\scriptstyle 0.07}$ & $0.98_{\scriptstyle 0.1}$ & $1.00_{\scriptstyle 0.00}$ & 0.07 & $0.70_{\scriptstyle 0.17}$ & $0.53_{\scriptstyle 0.24}$ & $1.00_{\scriptstyle 0.00}$ \\ 
				& S5 & 1.00 & $1.00_{\scriptstyle 0.01}$ & $1.00_{\scriptstyle 0.00}$ & $1.00_{\scriptstyle 0.00}$ & 0.96 & $1.00_{\scriptstyle 0.02}$ & $1.00_{\scriptstyle 0.03}$ & $1.00_{\scriptstyle 0.00}$ & 0.99 & $1.00_{\scriptstyle 0.01}$ & $1.00_{\scriptstyle 0.01}$ & $1.00_{\scriptstyle 0.00}$ & 0.96 & $1.00_{\scriptstyle 0.02}$ & $1.00_{\scriptstyle 0.03}$ & $1.00_{\scriptstyle 0.00}$ & 0.00 & $0.63_{\scriptstyle 0.15}$ & $0.44_{\scriptstyle 0.14}$ & $1.00_{\scriptstyle 0.00}$ \\
				& SCAD & 0.18 & $0.74_{\scriptstyle 0.19}$ & $1.00_{\scriptstyle 0.00}$ & $0.98_{\scriptstyle 0.02}$ & 0.41 & $0.86_{\scriptstyle 0.16}$ & $1.00_{\scriptstyle 0.00}$ & $0.99_{\scriptstyle 0.01}$ & 0.08 & $0.68_{\scriptstyle 0.18}$ & $1.00_{\scriptstyle 0.02}$ & $0.98_{\scriptstyle 0.02}$ & 0.01 & $0.52_{\scriptstyle 0.14}$ & $0.92_{\scriptstyle 0.15}$ & $0.97_{\scriptstyle 0.02}$ & 0.00 & $0.36_{\scriptstyle 0.15}$ & $0.40_{\scriptstyle 0.16}$ & $0.99_{\scriptstyle 0.01}$ \\ 
				& AdLa & 0.06 & $0.93_{\scriptstyle 0.10}$ & $1.00_{\scriptstyle 0.00}$ & $1.00_{\scriptstyle 0.00}$ & 0.41 & $0.89_{\scriptstyle 0.12}$ & $1.00_{\scriptstyle 0.01}$ & $0.99_{\scriptstyle 0.01}$ & 0.38 & $0.88_{\scriptstyle 0.13}$ & $0.99_{\scriptstyle 0.07}$ & $0.99_{\scriptstyle 0.01}$ & 0.19 & $0.79_{\scriptstyle 0.17}$ & $0.85_{\scriptstyle 0.21}$ & $1.00_{\scriptstyle 0.01}$ & 0.05 & $0.64_{\scriptstyle 0.19}$ & $0.55_{\scriptstyle 0.24}$ & $1.00_{\scriptstyle 0.00}$ \\ 
				& MCP & 0.51 & $0.88_{\scriptstyle 0.15}$ & $1.00_{\scriptstyle 0.00}$ & $1.00_{\scriptstyle 0.01}$ & 0.70 & $0.95_{\scriptstyle 0.05}$ & $1.00_{\scriptstyle 0.01}$ & $1.00_{\scriptstyle 0.00}$ & 0.40 & $0.86_{\scriptstyle 0.15}$ & $1.00_{\scriptstyle 0.01}$ & $0.99_{\scriptstyle 0.01}$ & 0.08 & $0.69_{\scriptstyle 0.18}$ & $0.89_{\scriptstyle 0.19}$ & $0.99_{\scriptstyle 0.01}$ & 0.00 & $0.42_{\scriptstyle 0.15}$ & $0.37_{\scriptstyle 0.15}$ & $0.99_{\scriptstyle 0.01}$ \\ 
				\hline 
				\multirow{4}{*}{\parbox{0.07 \linewidth}{\centering Set-2}} & SAVS & 0.61 & $0.94_{\scriptstyle 0.09}$ & $0.96_{\scriptstyle 0.08}$ & $1.00_{\scriptstyle 0.00}$ & 0.25 & $0.85_{\scriptstyle 0.14}$ & $0.82_{\scriptstyle 0.18}$ & $1.00_{\scriptstyle 0.00}$ & 0.36 & $0.87_{\scriptstyle 0.14}$ & $0.81_{\scriptstyle 0.21}$ & $0.99_{\scriptstyle 0.00}$ & 0.10 & $0.74_{\scriptstyle 0.18}$ & $0.60_{\scriptstyle 0.24}$ & $1.00_{\scriptstyle 0.00}$ & 0.00 & $0.52_{\scriptstyle 0.21}$ & $0.34_{\scriptstyle 0.19}$ & $1.00_{\scriptstyle 0.00}$ \\ 
				& S5 & 0.64 & $0.95_{\scriptstyle 0.07}$ & $0.92_{\scriptstyle 0.11}$ & $1.00_{\scriptstyle 0.00}$ & 0.15 & $0.83_{\scriptstyle 0.12}$ & $0.72_{\scriptstyle 0.18}$ & $1.00_{\scriptstyle 0.00}$ & 0.21 & $0.84_{\scriptstyle 0.12}$ & $0.73_{\scriptstyle 0.19}$ & $1.00_{\scriptstyle 0.00}$ & 0.02 & $0.73_{\scriptstyle 0.14}$ & $0.56_{\scriptstyle 0.17}$ & $1.00_{\scriptstyle 0.00}$ & 0.00 & $0.52_{\scriptstyle 0.18}$ & $0.34_{\scriptstyle 0.12}$ & $1.00_{\scriptstyle 0.00}$ \\
				& SCAD & 0.00 & $0.50_{\scriptstyle 0.12}$ & $1.00_{\scriptstyle 0.03}$ & $0.97_{\scriptstyle 0.02}$ & 0.01 & $0.62_{\scriptstyle 0.12}$ & $0.94_{\scriptstyle 0.11}$ & $0.98_{\scriptstyle 0.01}$ & 0.00 & $0.45_{\scriptstyle 0.11}$ & $0.91_{\scriptstyle 0.15}$ & $0.96_{\scriptstyle 0.02}$ & 0.00 & $0.37_{\scriptstyle 0.13}$ & $0.64_{\scriptstyle 0.20}$ & $0.97_{\scriptstyle 0.02}$ & 0.00 & $0.29_{\scriptstyle 0.14}$ & $0.31_{\scriptstyle 0.13}$ & $0.99_{\scriptstyle 0.01}$ \\ 
				& AdLa & 0.25 & $0.84_{\scriptstyle 0.14}$ & $0.98_{\scriptstyle 0.07}$ & $0.99_{\scriptstyle 0.01}$ & 0.08 & $0.75_{\scriptstyle 0.15}$ & $0.89_{\scriptstyle 0.14}$ & $0.99_{\scriptstyle 0.01}$ & 0.11 & $0.75_{\scriptstyle 0.16}$ & $0.82_{\scriptstyle 0.21}$ & $0.99_{\scriptstyle 0.01}$ & 0.06 & $0.66_{\scriptstyle 0.19}$ & $0.64_{\scriptstyle 0.24}$ & $1.00_{\scriptstyle 0.02}$ & 0.02 & $0.54_{\scriptstyle 0.22}$ & $0.42_{\scriptstyle 0.23}$ & $1.00_{\scriptstyle 0.00}$ \\ 
				& MCP & 0.09 & $0.74_{\scriptstyle 0.14}$ & $0.99_{\scriptstyle 0.04}$ & $0.99_{\scriptstyle 0.01}$ & 0.14 & $0.81_{\scriptstyle 0.13}$ & $0.88_{\scriptstyle 0.15}$ & $1.00_{\scriptstyle 0.00}$ & 0.01 & $0.63_{\scriptstyle 0.14}$ & $0.85_{\scriptstyle 0.19}$ & $0.99_{\scriptstyle 0.01}$ & 0.00 & $0.48_{\scriptstyle 0.15}$ & $0.55_{\scriptstyle 0.20}$ & $0.99_{\scriptstyle 0.01}$ & 0.00 & $0.35_{\scriptstyle 0.15}$ & $0.27_{\scriptstyle 0.12}$ & $0.99_{\scriptstyle 0.01}$ \\ 
				\hline 
	\end{tabular}}}
\end{table}

\begin{table}[H]
	\caption{Prop is the proportion of times true model being selected; for SCAD and MCP it corresponds to true model+intercept. Means and standard deviations(in subscript) for MCC, TPR and TNR over different methods are tabulated corresponding to $s_0 = 5$, $p = 500$ and $n = 200$. ``AdLa" is short for Adaptive LASSO}\label{tab-p500-n200}
	\centering
	{\resizebox{\textwidth}{!}{\begin{tabular}{|c c||c c c c|c c c c|c c c c|c c c c|c c c c|}\hline
				\multirow{2}{*}{\parbox{0.07 \linewidth}{$p$=500 \ $n$=200}} &  & \multicolumn{4}{|c|}{Independent} &  \multicolumn{4}{|c|}{Compound symmetry} & \multicolumn{4}{|c|}{\centering AR(1) with $\rho = 0.5$} & \multicolumn{4}{|c|}{\centering AR(1) with $\rho = 0.7$} & \multicolumn{4}{|c|}{\centering AR(1) with $\rho = 0.9$} \\ 
				& & Prop & MCC & TPR & TNR & Prop & MCC & TPR & TNR & Prop & MCC & TPR & TNR & Prop & MCC & TPR & TNR & Prop & MCC & TPR & TNR \\ 
				\hline 
				\multirow{4}{*}{\parbox{0.07 \linewidth}{\centering Set-1}} & SAVS & 0.85 & $0.98_{\scriptstyle 0.04}$ & $1.00_{\scriptstyle 0.00}$ & $1.00_{\scriptstyle 0.00}$ & 0.75 & $0.97_{\scriptstyle 0.05}$ & $1.00_{\scriptstyle 0.00}$ & $1.00_{\scriptstyle 0.00}$ & 0.88 & $0.99_{\scriptstyle 0.03}$ & $1.00_{\scriptstyle 0.00}$ & $1.00_{\scriptstyle 0.00}$ & 0.92 & $0.99_{\scriptstyle 0.03}$ & $1.00_{\scriptstyle 0.00}$ & $1.00_{\scriptstyle 0.00}$ & 0.61 & $0.91_{\scriptstyle 0.13}$ & $0.85_{\scriptstyle 0.21}$ & $1.00_{\scriptstyle 0.00}$ \\ 
				& S5 & 0.99 & $1.00_{\scriptstyle 0.01}$ & $1.00_{\scriptstyle 0.00}$ & $1.00_{\scriptstyle 0.00}$ & 0.99 & $1.00_{\scriptstyle 0.01}$ & $1.00_{\scriptstyle 0.00}$ & $1.00_{\scriptstyle 0.00}$ & 0.98 & $1.00_{\scriptstyle 0.00}$ & $1.00_{\scriptstyle 0.00}$ & $1.00_{\scriptstyle 0.00}$ & 1.00 & $1.00_{\scriptstyle 0.00}$ & $1.00_{\scriptstyle 0.00}$ & $1.00_{\scriptstyle 0.00}$ & 0.13 & $0.79_{\scriptstyle 0.13}$ & $0.65_{\scriptstyle 0.18}$ & $1.00_{\scriptstyle 0.00}$ \\
				& SCAD & 0.56 & $0.86_{\scriptstyle 0.20}$ & $1.00_{\scriptstyle 0.00}$ & $0.99_{\scriptstyle 0.02}$ & 0.69 & $0.92_{\scriptstyle 0.14}$ & $1.00_{\scriptstyle 0.00}$ & $1.00_{\scriptstyle 0.01}$ & 0.56 & $0.87_{\scriptstyle 0.19}$ & $1.00_{\scriptstyle 0.00}$ & $0.99_{\scriptstyle 0.01}$ & 0.35 & $0.82_{\scriptstyle 0.18}$ & $1.00_{\scriptstyle 0.02}$ & $0.99_{\scriptstyle 0.01}$ & 0.00 & $0.47_{\scriptstyle 0.16}$ & $0.62_{\scriptstyle 0.20}$ & $0.98_{\scriptstyle 0.02}$ \\ 
				& AdLa & 0.7 & $0.95_{\scriptstyle 0.09}$ & $1.00_{\scriptstyle 0.00}$ & $0.99_{\scriptstyle 0.00}$ & 0.61 & $0.94_{\scriptstyle 0.10}$ & $1.00_{\scriptstyle 0.00}$ & $0.99_{\scriptstyle 0.00}$ & 0.61 & $0.94_{\scriptstyle 0.09}$ & $1.00_{\scriptstyle 0.00}$ & $0.99_{\scriptstyle 0.00}$ & 0.45 & $0.90_{\scriptstyle 0.11}$ & $0.99_{\scriptstyle 0.03}$ & $0.99_{\scriptstyle 0.00}$ & 0.13 & $0.75_{\scriptstyle 0.17}$ & $0.76_{\scriptstyle 0.22}$ & $0.99_{\scriptstyle 0.00}$ \\ 
				& MCP & 0.67 & $0.91_{\scriptstyle 0.15}$ & $1.00_{\scriptstyle 0.00}$ & $1.00_{\scriptstyle 0.01}$ & 0.77 & $0.96_{\scriptstyle 0.08}$ & $1.00_{\scriptstyle 0.00}$ & $1.00_{\scriptstyle 0.00}$ & 0.67 & $0.92_{\scriptstyle 0.14}$ & $1.00_{\scriptstyle 0.00}$ & $1.00_{\scriptstyle 0.01}$ & 0.60 & $0.91_{\scriptstyle 0.14}$ & $1.00_{\scriptstyle 0.01}$ & $1.00_{\scriptstyle 0.01}$ & 0.00 & $0.54_{\scriptstyle 0.16}$ & $0.57_{\scriptstyle 0.18}$ & $0.99_{\scriptstyle 0.01}$ \\ 
				\hline 
				\multirow{4}{*}{\parbox{0.07 \linewidth}{\centering Set-2}} & SAVS & 0.85 & $0.99_{\scriptstyle 0.04}$ & $1.00_{\scriptstyle 0.01}$ & $1.00_{\scriptstyle 0.00}$ & 0.70 & $0.96_{\scriptstyle 0.06}$ & $0.98_{\scriptstyle 0.06}$ & $1.00_{\scriptstyle 0.00}$ & 0.86 & $0.99_{\scriptstyle 0.04}$ & $1.00_{\scriptstyle 0.04}$ & $1.00_{\scriptstyle 0.00}$ & 0.62 & $0.93_{\scriptstyle 0.10}$ & $0.89_{\scriptstyle 0.18}$ & $0.99_{\scriptstyle 0.00}$ & 0.04 & $0.68_{\scriptstyle 0.17}$ & $0.50_{\scriptstyle 0.22}$ & $1.00_{\scriptstyle 0.00}$ \\ 
				& S5 & 0.97 & $1.00_{\scriptstyle 0.02}$ & $1.00_{\scriptstyle 0.02}$ & $1.00_{\scriptstyle 0.00}$ & 0.82 & $0.98_{\scriptstyle 0.05}$ & $0.97_{\scriptstyle 0.08}$ & $1.00_{\scriptstyle 0.00}$ & 0.93 & $0.99_{\scriptstyle 0.04}$ & $0.99_{\scriptstyle 0.06}$ & $1.00_{\scriptstyle 0.00}$ & 0.59 & $0.92_{\scriptstyle 0.11}$ & $0.87_{\scriptstyle 0.18}$ & $1.00_{\scriptstyle 0.00}$ & 0.00 & $0.64_{\scriptstyle 0.15}$ & $0.45_{\scriptstyle 0.14}$ & $1.00_{\scriptstyle 0.00}$ \\
				& SCAD & 0.04 & $0.64_{\scriptstyle 0.18}$ & $1.00_{\scriptstyle 0.00}$ & $0.98_{\scriptstyle 0.02}$ & 0.08 & $0.72_{\scriptstyle 0.15}$ & $1.00_{\scriptstyle 0.03}$ & $0.99_{\scriptstyle 0.01}$ & 0.01 & $0.55_{\scriptstyle 0.15}$ & $1.00_{\scriptstyle 0.02}$ & $0.97_{\scriptstyle 0.02}$ & 0.00 & $0.47_{\scriptstyle 0.13}$ & $0.91_{\scriptstyle 0.15}$ & $0.96_{\scriptstyle 0.02}$ & 0.00 & $0.36_{\scriptstyle 0.15}$ & $0.43_{\scriptstyle 0.16}$ & $0.98_{\scriptstyle 0.02}$ \\ 
				& AdLa & 0.52 & $0.92_{\scriptstyle 0.11}$ & $0.99_{\scriptstyle 0.00}$ & $0.99_{\scriptstyle 0.00}$ & 0.30 & $0.87_{\scriptstyle 0.13}$ & $0.99_{\scriptstyle 0.04}$ & $1.00_{\scriptstyle 0.01}$ & 0.33 & $0.87_{\scriptstyle 0.13}$ & $0.99_{\scriptstyle 0.05}$ & $0.99_{\scriptstyle 0.01}$ & 0.16 & $0.80_{\scriptstyle 0.15}$ & $0.85_{\scriptstyle 0.20}$ & $0.99_{\scriptstyle 0.01}$ & 0.06 & $0.64_{\scriptstyle 0.2}$ & $0.54_{\scriptstyle 0.24}$ & $0.99_{\scriptstyle 0.00}$ \\ 
				& MCP & 0.34 & $0.83_{\scriptstyle 0.17}$ & $1.00_{\scriptstyle 0.00}$ & $0.99_{\scriptstyle 0.01}$ & 0.46 & $0.91_{\scriptstyle 0.10}$ & $0.99_{\scriptstyle 0.04}$ & $1.00_{\scriptstyle 0.00}$ & 0.17 & $0.78_{\scriptstyle 0.16}$ & $1.00_{\scriptstyle 0.03}$ & $0.99_{\scriptstyle 0.01}$ & 0.02 & $0.63_{\scriptstyle 0.14}$ & $0.88_{\scriptstyle 0.17}$ & $0.99_{\scriptstyle 0.01}$ & 0.00 & $0.43_{\scriptstyle 0.15}$ & $0.40_{\scriptstyle 0.16}$ & $0.99_{\scriptstyle 0.01}$ \\ 
				\hline 
	\end{tabular}}} 
\end{table}

\begin{table}[H]
	\caption{Prop is the proportion of times true model being selected; for SCAD and MCP it corresponds to true model+intercept. Means and standard deviations(in subscript) for MCC, TPR and TNR over different methods are tabulated corresponding to $s_0 = 5$, $p = 1000$ and $n = 100$. ``AdLa" is short for Adaptive LASSO}\label{tab-p1000-n100}
	\centering
	{\resizebox{\textwidth}{!}{\begin{tabular}{|c c||c c c c|c c c c|c c c c|c c c c|c c c c|}\hline
				\multirow{2}{*}{\parbox{0.07 \linewidth}{$p$=1000 \ $n$=100}} &  & \multicolumn{4}{|c|}{Independent} &  \multicolumn{4}{|c|}{Compound symmetry} & \multicolumn{4}{|c|}{\centering AR(1) with $\rho = 0.5$} & \multicolumn{4}{|c|}{\centering AR(1) with $\rho = 0.7$} & \multicolumn{4}{|c|}{\centering AR(1) with $\rho = 0.9$} \\ 
				& & Prop & MCC & TPR & TNR & Prop & MCC & TPR & TNR & Prop & MCC & TPR & TNR & Prop & MCC & TPR & TNR & Prop & MCC & TPR & TNR \\ 
				\hline 
				\multirow{4}{*}{\parbox{0.07 \linewidth}{\centering Set-1}} & SAVS & 0.82 & $0.97_{\scriptstyle 0.08}$ & $1.00_{\scriptstyle 0.00}$ & $1.00_{\scriptstyle 0.00}$ & 0.79 & $0.96_{\scriptstyle 0.10}$ & $1.00_{\scriptstyle 0.02}$ & $1.00_{\scriptstyle 0.00}$ & 0.86 & $0.98_{\scriptstyle 0.07}$ & $1.00_{\scriptstyle 0.02}$ & $1.00_{\scriptstyle 0.00}$ & 0.73 & $0.95_{\scriptstyle 0.11}$ & $0.94_{\scriptstyle 0.15}$ & $1.00_{\scriptstyle 0.00}$ & 0.03 & $0.67_{\scriptstyle 0.16}$ & $0.48_{\scriptstyle 0.21}$ & $1.00_{\scriptstyle 0.00}$ \\
				& S5 & 1.00 & $1.00_{\scriptstyle 0.00}$ & $1.00_{\scriptstyle 0.00}$ & $1.00_{\scriptstyle 0.00}$ & 0.94 & $0.99_{\scriptstyle 0.03}$ & $0.99_{\scriptstyle 0.06}$ & $1.00_{\scriptstyle 0.00}$ & 0.97 & $0.99_{\scriptstyle 0.05}$ & $0.99_{\scriptstyle 0.06}$ & $1.00_{\scriptstyle 0.00}$ & 0.69 & $0.93_{\scriptstyle 0.12}$ & $0.88_{\scriptstyle 0.20}$ & $1.00_{\scriptstyle 0.00}$ & 0.00 & $0.59_{\scriptstyle 0.16}$ & $0.40_{\scriptstyle 0.14}$ & $1.00_{\scriptstyle 0.00}$ \\ 
				& SCAD & 0.11 & $0.68_{\scriptstyle 0.19}$ & $1.00_{\scriptstyle 0.00}$ & $0.99_{\scriptstyle 0.01}$ & 0.31 & $0.81_{\scriptstyle 0.17}$ & $1.00_{\scriptstyle 0.01}$ & $1.00_{\scriptstyle 0.00}$ & 0.03 & $0.60_{\scriptstyle 0.17}$ & $1.00_{\scriptstyle 0.02}$ & $0.99_{\scriptstyle 0.01}$ & 0.00 & $0.44_{\scriptstyle 0.14}$ & $0.83_{\scriptstyle 0.20}$ & $0.98_{\scriptstyle 0.01}$ & 0.00 & $0.32_{\scriptstyle 0.15}$ & $0.36_{\scriptstyle 0.14}$ & $0.99_{\scriptstyle 0.01}$ \\ 
				& AdLa & 0.60 & $0.94_{\scriptstyle 0.09}$ & $1.00_{\scriptstyle 0.00}$ & $1.00_{\scriptstyle 0.00}$ & 0.34 & $0.88_{\scriptstyle 0.12}$ & $1.00_{\scriptstyle 0.02}$ & $1.00_{\scriptstyle 0.00}$ & 0.35 & $0.86_{\scriptstyle 0.15}$ & $0.95_{\scriptstyle 0.13}$ & $1.00_{\scriptstyle 0.00}$ & 0.14 & $0.74_{\scriptstyle 0.19}$ & $0.76_{\scriptstyle 0.23}$ & $1.00_{\scriptstyle 0.00}$ & 0.05 & $0.62_{\scriptstyle 0.20}$ & $0.51_{\scriptstyle 0.24}$ & $1.00_{\scriptstyle 0.00}$ \\ 
				& MCP & 0.46 & $0.87_{\scriptstyle 0.16}$ & $1.00_{\scriptstyle 0.00}$ & $1.00_{\scriptstyle 0.00}$ & 0.67 & $0.95_{\scriptstyle 0.09}$ & $1.00_{\scriptstyle 0.02}$ & $1.00_{\scriptstyle 0.00}$ & 0.32 & $0.82_{\scriptstyle 0.17}$ & $0.99_{\scriptstyle 0.05}$ & $1.00_{\scriptstyle 0.00}$ & 0.05 & $0.61_{\scriptstyle 0.19}$ & $0.77_{\scriptstyle 0.23}$ & $1.00_{\scriptstyle 0.00}$ & 0.00 & $0.39_{\scriptstyle 0.15}$ & $0.33_{\scriptstyle 0.14}$ & $1.00_{\scriptstyle 0.00}$ \\ 
				\hline 
				\multirow{4}{*}{\parbox{0.07 \linewidth}{\centering Set-2}} & SAVS & 0.52 & $0.93_{\scriptstyle 0.09}$ & $0.92_{\scriptstyle 0.12}$ & $1.00_{\scriptstyle 0.00}$ & 0.12 & $0.80_{\scriptstyle 0.15}$ & $0.73_{\scriptstyle 0.2}$ & $1.00_{\scriptstyle 0.00}$ & 0.23 & $0.83_{\scriptstyle 0.14}$ & $0.74_{\scriptstyle 0.22}$ & $1.00_{\scriptstyle 0.00}$ & 0.03 & $0.70_{\scriptstyle 0.17}$ & $0.54_{\scriptstyle 0.21}$ & $1.00_{\scriptstyle 0.00}$ & 0.00 & $0.48_{\scriptstyle 0.2}$ & $0.28_{\scriptstyle 0.16}$ & $1.00_{\scriptstyle 0.00}$ \\ 
				& S5 & 0.52 & $0.94_{\scriptstyle 0.07}$ & $0.89_{\scriptstyle 0.13}$ & $1.00_{\scriptstyle 0.00}$ & 0.08 & $0.79_{\scriptstyle 0.13}$ & $0.66_{\scriptstyle 0.18}$ & $1.00_{\scriptstyle 0.00}$ & 0.12 & $0.81_{\scriptstyle 0.13}$ & $0.68_{\scriptstyle 0.19}$ & $1.00_{\scriptstyle 0.00}$ & 0.01 & $0.71_{\scriptstyle 0.14}$ & $0.53_{\scriptstyle 0.17}$ & $1.00_{\scriptstyle 0.00}$ & 0.00 & $0.48_{\scriptstyle 0.18}$ & $0.31_{\scriptstyle 0.12}$ & $1.00_{\scriptstyle 0.00}$ \\
				& SCAD & 0.00 & $0.44_{\scriptstyle 0.10}$ & $0.99_{\scriptstyle 0.03}$ & $0.98_{\scriptstyle 0.01}$ & 0.00 & $0.57_{\scriptstyle 0.12}$ & $0.92_{\scriptstyle 0.12}$ & $0.99_{\scriptstyle 0.00}$ & 0.00 & $0.40_{\scriptstyle 0.11}$ & $0.85_{\scriptstyle 0.18}$ & $0.98_{\scriptstyle 0.01}$ & 0.00 & $0.32_{\scriptstyle 0.14}$ & $0.57_{\scriptstyle 0.20}$ & $0.98_{\scriptstyle 0.01}$ & 0.00 & $0.28_{\scriptstyle 0.15}$ & $0.28_{\scriptstyle 0.12}$ & $0.99_{\scriptstyle 0.01}$ \\ 
				& AdLa & 0.21 & $0.84_{\scriptstyle 0.13}$ & $0.96_{\scriptstyle 0.08}$ & $1.00_{\scriptstyle 0.00}$ & 0.04 & $0.71_{\scriptstyle 0.15}$ & $0.84_{\scriptstyle 0.15}$ & $0.99_{\scriptstyle 0.00}$ & 0.08 & $0.71_{\scriptstyle 0.18}$ & $0.75_{\scriptstyle 0.23}$ & $1.00_{\scriptstyle 0.00}$ & 0.05 & $0.62_{\scriptstyle 0.21}$ & $0.58_{\scriptstyle 0.26}$ & $1.00_{\scriptstyle 0.00}$ & 0.01 & $0.53_{\scriptstyle 0.22}$ & $0.41_{\scriptstyle 0.23}$ & $1.00_{\scriptstyle 0.00}$ \\ 
				& MCP & 0.04 & $0.70_{\scriptstyle 0.14}$ & $0.99_{\scriptstyle 0.05}$ & $0.99_{\scriptstyle 0.00}$ & 0.10 & $0.78_{\scriptstyle 0.15}$ & $0.84_{\scriptstyle 0.17}$ & $1.00_{\scriptstyle 0.00}$ & 0.01 & $0.57_{\scriptstyle 0.15}$ & $0.78_{\scriptstyle 0.20}$ & $0.99_{\scriptstyle 0.00}$ & 0.00 & $0.44_{\scriptstyle 0.15}$ & $0.49_{\scriptstyle 0.18}$ & $0.99_{\scriptstyle 0.00}$ & 0.00 & $0.35_{\scriptstyle 0.15}$ & $0.25_{\scriptstyle 0.11}$ & $1.00_{\scriptstyle 0.00}$ \\ 
				\hline 
	\end{tabular}}}
\end{table}

\begin{table}[H]
	\caption{Prop is the proportion of times true model being selected; for SCAD and MCP it corresponds to true model+intercept. Means and standard deviations(in subscript) for MCC, TPR and TNR over different methods are tabulated corresponding to $s_0 = 5$, $p = 1000$ and $n = 200$. ``AdLa" is short for Adaptive LASSO}\label{tab-p1000-n200}
	\centering
	{\resizebox{\textwidth}{!}{\begin{tabular}{|c c||c c c c|c c c c|c c c c|c c c c|c c c c|}\hline
				\multirow{2}{*}{\parbox{0.07 \linewidth}{$p$=1000 \ $n$=200}} &  & \multicolumn{4}{|c|}{Independent} &  \multicolumn{4}{|c|}{Compound symmetry} & \multicolumn{4}{|c|}{\centering AR(1) with $\rho = 0.5$} & \multicolumn{4}{|c|}{\centering AR(1) with $\rho = 0.7$} & \multicolumn{4}{|c|}{\centering AR(1) with $\rho = 0.9$} \\ 
				& & Prop & MCC & TPR & TNR & Prop & MCC & TPR & TNR & Prop & MCC & TPR & TNR & Prop & MCC & TPR & TNR & Prop & MCC & TPR & TNR \\ 
				\hline 
				\multirow{4}{*}{\parbox{0.07 \linewidth}{\centering Set-1}} & SAVS & 0.95 & $1.00_{\scriptstyle 0.02}$ & $1.00_{\scriptstyle 0.00}$ & $1.00_{\scriptstyle 0.00}$ & 0.95 & $0.99_{\scriptstyle 0.04}$ & $1.00_{\scriptstyle 0.00}$ & $1.00_{\scriptstyle 0.00}$ & 0.97 & $1.00_{\scriptstyle 0.02}$ & $1.00_{\scriptstyle 0.00}$ & $1.00_{\scriptstyle 0.00}$ & 0.98 & $1.00_{\scriptstyle 0.01}$ & $1.00_{\scriptstyle 0.00}$ & $1.00_{\scriptstyle 0.00}$ & 0.54 & $0.90_{\scriptstyle 0.13}$ & $0.82_{\scriptstyle 0.21}$ & $1.00_{\scriptstyle 0.00}$ \\ 
				& S5 & 1.00 & $1.00_{\scriptstyle 0.00}$ & $1.00_{\scriptstyle 0.00}$ & $1.00_{\scriptstyle 0.00}$ & 0.99 & $1.00_{\scriptstyle 0.01}$ & $1.00_{\scriptstyle 0.00}$ & $1.00_{\scriptstyle 0.00}$ & 0.99 & $1.00_{\scriptstyle 0.01}$ & $1.00_{\scriptstyle 0.00}$ & $1.00_{\scriptstyle 0.00}$ & 1.00 & $1.00_{\scriptstyle 0.01}$ & $1.00_{\scriptstyle 0.00}$ & $1.00_{\scriptstyle 0.00}$ & 0.02 & $0.74_{\scriptstyle 0.12}$ & $0.56_{\scriptstyle 0.16}$ & $1.00_{\scriptstyle 0.00}$ \\
				& SCAD & 0.48 & $0.82_{\scriptstyle 0.23}$ & $1.00_{\scriptstyle 0.00}$ & $0.99_{\scriptstyle 0.01}$ & 0.64 & $0.90_{\scriptstyle 0.16}$ & $1.00_{\scriptstyle 0.00}$ & $1.00_{\scriptstyle 0.00}$ & 0.47 & $0.83_{\scriptstyle 0.22}$ & $1.00_{\scriptstyle 0.00}$ & $0.99_{\scriptstyle 0.01}$ & 0.24 & $0.75_{\scriptstyle 0.21}$ & $1.00_{\scriptstyle 0.02}$ & $0.99_{\scriptstyle 0.01}$ & 0.00 & $0.41_{\scriptstyle 0.18}$ & $0.55_{\scriptstyle 0.18}$ & $0.99_{\scriptstyle 0.01}$ \\ 
				& AdLa & 0.71 & $0.95_{\scriptstyle 0.08}$ & $1.00_{\scriptstyle 0.00}$ & $0.99_{\scriptstyle 0.00}$ & 0.64 & $0.94_{\scriptstyle 0.01}$ & $1.00_{\scriptstyle 0.00}$ & $0.99_{\scriptstyle 0.00}$ & 0.62 & $0.94_{\scriptstyle 0.01}$ & $1.00_{\scriptstyle 0.00}$ & $1.00_{\scriptstyle 0.002}$ & 0.40 & $0.88_{\scriptstyle 0.13}$ & $0.98_{\scriptstyle 0.06}$ & $1.00_{\scriptstyle 0.002}$ & 0.10 & $0.71_{\scriptstyle 0.19}$ & $0.65_{\scriptstyle 0.24}$ & $1.00_{\scriptstyle 0.002}$ \\ 
				& MCP & 0.64 & $0.90_{\scriptstyle 0.17}$ & $1.00_{\scriptstyle 0.00}$ & $1.00_{\scriptstyle 0.00}$ & 0.76 & $0.96_{\scriptstyle 0.08}$ & $1.00_{\scriptstyle 0.00}$ & $1.00_{\scriptstyle 0.00}$ & 0.64 & $0.90_{\scriptstyle 0.16}$ & $1.00_{\scriptstyle 0.00}$ & $1.00_{\scriptstyle 0.00}$ & 0.55 & $0.89_{\scriptstyle 0.15}$ & $1.00_{\scriptstyle 0.02}$ & $1.00_{\scriptstyle 0.00}$ & 0.00 & $0.50_{\scriptstyle 0.18}$ & $0.50_{\scriptstyle 0.16}$ & $1.00_{\scriptstyle 0.01}$ \\ 
				\hline 
				\multirow{4}{*}{\parbox{0.07 \linewidth}{\centering Set-2}} & SAVS & 0.93 & $0.99_{\scriptstyle 0.03}$ & $1.00_{\scriptstyle 0.03}$ & $1.00_{\scriptstyle 0.00}$ & 0.73 & $0.97_{\scriptstyle 0.06}$ & $0.95_{\scriptstyle 0.09}$ & $1.00_{\scriptstyle 0.00}$ & 0.87 & $0.98_{\scriptstyle 0.05}$ & $0.97_{\scriptstyle 0.09}$ & $1.00_{\scriptstyle 0.00}$ & 0.47 & $0.9_{\scriptstyle 0.11}$ & $0.82_{\scriptstyle 0.19}$ & $1.00_{\scriptstyle 0.00}$ & 0.00 & $0.65_{\scriptstyle 0.14}$ & $0.45_{\scriptstyle 0.17}$ & $1.00_{\scriptstyle 0.00}$ \\ 
				& S5 & 0.97 & $1.00_{\scriptstyle 0.02}$ & $1.00_{\scriptstyle 0.02}$ & $1.00_{\scriptstyle 0.00}$ & 0.79 & $0.97_{\scriptstyle 0.06}$ & $0.96_{\scriptstyle 0.09}$ & $1.00_{\scriptstyle 0.00}$ & 0.90 & $0.99_{\scriptstyle 0.05}$ & $0.98_{\scriptstyle 0.08}$ & $1.00_{\scriptstyle 0.00}$ & 0.47 & $0.90_{\scriptstyle 0.11}$ & $0.82_{\scriptstyle 0.19}$ & $1.00_{\scriptstyle 0.00}$ & 0.00 & $0.61_{\scriptstyle 0.16}$ & $0.42_{\scriptstyle 0.14}$ & $1.00_{\scriptstyle 0.00}$ \\
				& SCAD & 0.02 & $0.56_{\scriptstyle 0.17}$ & $1.00_{\scriptstyle 0.00}$ & $0.98_{\scriptstyle 0.01}$ & 0.06 & $0.67_{\scriptstyle 0.16}$ & $1.00_{\scriptstyle 0.03}$ & $0.99_{\scriptstyle 0.01}$ & 0.01 & $0.47_{\scriptstyle 0.15}$ & $0.99_{\scriptstyle 0.04}$ & $0.98_{\scriptstyle 0.01}$ & 0.00 & $0.40_{\scriptstyle 0.13}$ & $0.85_{\scriptstyle 0.17}$ & $0.98_{\scriptstyle 0.02}$ & 0.00 & $0.32_{\scriptstyle 0.15}$ & $0.40_{\scriptstyle 0.15}$ & $0.99_{\scriptstyle 0.01}$ \\ 
				& AdLa & 0.48 & $0.91_{\scriptstyle 0.11}$ & $0.99_{\scriptstyle 0.01}$ & $0.99_{\scriptstyle 0.00}$ & 0.28 & $0.86_{\scriptstyle 0.13}$ & $0.98_{\scriptstyle 0.06}$ & $0.99_{\scriptstyle 0.00}$ & 0.28 & $0.85_{\scriptstyle 0.14}$ & $0.97_{\scriptstyle 0.09}$ & $0.99_{\scriptstyle 0.00}$ & 0.11 & $0.76_{\scriptstyle 0.16}$ & $0.78_{\scriptstyle 0.21}$ & $0.99_{\scriptstyle 0.00}$ & 0.052 & $0.62_{\scriptstyle 0.20}$ & $0.50_{\scriptstyle 0.23}$ & $1.00_{\scriptstyle 0.002}$ \\ 
				& MCP & 0.26 & $0.80_{\scriptstyle 0.18}$ & $1.00_{\scriptstyle 0.01}$ & $1.00_{\scriptstyle 0.01}$ & 0.41 & $0.90_{\scriptstyle 0.11}$ & $0.99_{\scriptstyle 0.05}$ & $1.00_{\scriptstyle 0.00}$ & 0.10 & $0.72_{\scriptstyle 0.17}$ & $0.99_{\scriptstyle 0.05}$ & $0.99_{\scriptstyle 0.01}$ & 0.00 & $0.58_{\scriptstyle 0.15}$ & $0.81_{\scriptstyle 0.19}$ & $0.99_{\scriptstyle 0.01}$ & 0.00 & $0.39_{\scriptstyle 0.15}$ & $0.37_{\scriptstyle 0.16}$ & $1.00_{\scriptstyle 0.00}$ \\ 
				\hline 
	\end{tabular}}}
\end{table}

\begin{table}[H]
	\caption{Prop is the proportion of times true model being selected; for SCAD and MCP it corresponds to true model+intercept. Means and standard deviations(in subscript) for MCC, TPR and TNR over different methods are tabulated corresponding to $s_0 = 5$, $p = 5000$ and $n = 100$. ``AdLa" is short for Adaptive LASSO}\label{tab-p5000-n100}
	\centering
	{\resizebox{\textwidth}{!}{\begin{tabular}{|c c||c c c c|c c c c|c c c c|c c c c|c c c c|}\hline
				\multirow{2}{*}{\parbox{0.07 \linewidth}{$p$=5000 \ $n$=100}} &  & \multicolumn{4}{|c|}{Independent} &  \multicolumn{4}{|c|}{Compound symmetry} & \multicolumn{4}{|c|}{\centering AR(1) with $\rho = 0.5$} & \multicolumn{4}{|c|}{\centering AR(1) with $\rho = 0.7$} & \multicolumn{4}{|c|}{\centering AR(1) with $\rho = 0.9$} \\ 
				& & Prop & MCC & TPR & TNR & Prop & MCC & TPR & TNR & Prop & MCC & TPR & TNR & Prop & MCC & TPR & TNR & Prop & MCC & TPR & TNR \\ 
				\hline 
				\multirow{4}{*}{\parbox{0.07 \linewidth}{\centering Set-1}} & SAVS & 0.75 & $0.92_{\scriptstyle 0.17}$ & $0.91_{\scriptstyle 0.20}$ & $1.00_{\scriptstyle 0.00}$ & 0.79 & $0.96_{\scriptstyle 0.10}$ & $0.98_{\scriptstyle 0.10}$ & $1.00_{\scriptstyle 0.00}$ & 0.75 & $0.92_{\scriptstyle 0.17}$ & $0.91_{\scriptstyle 0.20}$ & $1.00_{\scriptstyle 0.00}$ & 0.39 & $0.82_{\scriptstyle 0.20}$ & $0.73_{\scriptstyle 0.26}$ & $1.00_{\scriptstyle 0.00}$ & 0.00 & $0.60_{\scriptstyle 0.16}$ & $0.40_{\scriptstyle 0.18}$ & $1.00_{\scriptstyle 0.00}$ \\ 
				& S5 & 1.00 & $1.00_{\scriptstyle 0.01}$ & $1.00_{\scriptstyle 0.01}$ & $1.00_{\scriptstyle 0.00}$ & 0.88 & $0.98_{\scriptstyle 0.07}$ & $0.96_{\scriptstyle 0.11}$ & $1.00_{\scriptstyle 0.00}$ & 0.92 & $0.98_{\scriptstyle 0.07}$ & $0.97_{\scriptstyle 0.11}$ & $1.00_{\scriptstyle 0.00}$ & 0.32 & $0.81_{\scriptstyle 0.17}$ & $0.69_{\scriptstyle 0.25}$ & $1.00_{\scriptstyle 0.00}$ & 0.00 & $0.51_{\scriptstyle 0.18}$ & $0.33_{\scriptstyle 0.13}$ & $1.00_{\scriptstyle 0.00}$ \\
				& SCAD & 0.02 & $0.55_{\scriptstyle 0.17}$ & $1.00_{\scriptstyle 0.00}$ & $1.00_{\scriptstyle 0.00}$ & 0.16 & $0.71_{\scriptstyle 0.19}$ & $1.00_{\scriptstyle 0.03}$ & $1.00_{\scriptstyle 0.00}$ & 0.00 & $0.44_{\scriptstyle 0.13}$ & $0.96_{\scriptstyle 0.12}$ & $1.00_{\scriptstyle 0.00}$ & 0.00 & $0.32_{\scriptstyle 0.14}$ & $0.64_{\scriptstyle 0.22}$ & $1.00_{\scriptstyle 0.00}$ & 0.00 & $0.29_{\scriptstyle 0.16}$ & $0.31_{\scriptstyle 0.12}$ & $1.00_{\scriptstyle 0.00}$ \\ 
				& AdLa & 0.26 & $0.79_{\scriptstyle 0.19}$ & $0.82_{\scriptstyle 0.22}$ & $1.00_{\scriptstyle 0.00}$ & 0.23 & $0.85_{\scriptstyle 0.13}$ & $0.98_{\scriptstyle 0.07}$ & $1.00_{\scriptstyle 0.00}$ & 0.26 & $0.79_{\scriptstyle 0.19}$ & $0.82_{\scriptstyle 0.23}$ & $1.00_{\scriptstyle 0.00}$ & 0.10 & $0.69_{\scriptstyle 0.21}$ & $0.63_{\scriptstyle 0.26}$ & $1.00_{\scriptstyle 0.00}$ & 0.05 & $0.61_{\scriptstyle 0.21}$ & $0.47_{\scriptstyle 0.24}$ & $1.00_{\scriptstyle 0.00}$ \\ 
				& MCP & 0.34 & $0.83_{\scriptstyle 0.16}$ & $1.00_{\scriptstyle 0.00}$ & $1.00_{\scriptstyle 0.00}$ & 0.58 & $0.93_{\scriptstyle 0.10}$ & $0.99_{\scriptstyle 0.05}$ & $1.00_{\scriptstyle 0.00}$ & 0.17 & $0.72_{\scriptstyle 0.20}$ & $0.93_{\scriptstyle 0.17}$ & $1.00_{\scriptstyle 0.00}$ & 0.01 & $0.48_{\scriptstyle 0.18}$ & $0.56_{\scriptstyle 0.22}$ & $1.00_{\scriptstyle 0.00}$ & 0.00 & $0.37_{\scriptstyle 0.15}$ & $0.28_{\scriptstyle 0.12}$ & $1.00_{\scriptstyle 0.00}$ \\ 
				\hline 
				\multirow{4}{*}{\parbox{0.07 \linewidth}{\centering Set-2}} & SAVS & 0.41 & $0.91_{\scriptstyle 0.11}$ & $0.86_{\scriptstyle 0.15}$ & $1.00_{\scriptstyle 0.00}$ & 0.05 & $0.71_{\scriptstyle 0.20}$ & $0.60_{\scriptstyle 0.22}$ & $1.00_{\scriptstyle 0.00}$ & 0.09 & $0.73_{\scriptstyle 0.19}$ & $0.60_{\scriptstyle 0.23}$ & $1.00_{\scriptstyle 0.00}$ & 0.00 & $0.62_{\scriptstyle 0.20}$ & $0.45_{\scriptstyle 0.21}$ & $1.00_{\scriptstyle 0.00}$ & 0.00 & $0.47_{\scriptstyle 0.2}$ & $0.27_{\scriptstyle 0.15}$ & $1.00_{\scriptstyle 0.00}$ \\ 
				& S5 & 0.28 & $0.89_{\scriptstyle 0.09}$ & $0.80_{\scriptstyle 0.16}$ & $1.00_{\scriptstyle 0.00}$ & 0.01 & $0.70_{\scriptstyle 0.16}$ & $0.54_{\scriptstyle 0.18}$ & $1.00_{\scriptstyle 0.00}$ & 0.01 & $0.74_{\scriptstyle 0.14}$ & $0.58_{\scriptstyle 0.18}$ & $1.00_{\scriptstyle 0.00}$ & 0.00 & $0.63_{\scriptstyle 0.16}$ & $0.44_{\scriptstyle 0.15}$ & $1.00_{\scriptstyle 0.00}$ & 0.00 & $0.42_{\scriptstyle 0.19}$ & $0.27_{\scriptstyle 0.12}$ & $1.00_{\scriptstyle 0.00}$ \\
				& SCAD & 0.00 & $0.36_{\scriptstyle 0.05}$ & $0.98_{\scriptstyle 0.05}$ & $0.99_{\scriptstyle 0.00}$ & 0.00 & $0.45_{\scriptstyle 0.11}$ & $0.84_{\scriptstyle 0.17}$ & $1.00_{\scriptstyle 0.00}$ & 0.00 & $0.31_{\scriptstyle 0.12}$ & $0.71_{\scriptstyle 0.21}$ & $0.99_{\scriptstyle 0.00}$ & 0.00 & $0.26_{\scriptstyle 0.15}$ & $0.46_{\scriptstyle 0.18}$ & $1.00_{\scriptstyle 0.00}$ & 0.00 & $0.24_{\scriptstyle 0.16}$ & $0.26_{\scriptstyle 0.13}$ & $1.00_{\scriptstyle 0.00}$ \\ 
				& AdLa & 0.12 & $0.80_{\scriptstyle 0.14}$ & $0.90_{\scriptstyle 0.13}$ & $1.00_{\scriptstyle 0.00}$ & 0.01 & $0.62_{\scriptstyle 0.17}$ & $0.71_{\scriptstyle 0.20}$ & $1.00_{\scriptstyle 0.00}$ & 0.06 & $0.66_{\scriptstyle 0.21}$ & $0.63_{\scriptstyle 0.25}$ & $1.00_{\scriptstyle 0.00}$ & 0.04 & $0.58_{\scriptstyle 0.23}$ & $0.49_{\scriptstyle 0.25}$ & $1.00_{\scriptstyle 0.00}$ & 0.01 & $0.51_{\scriptstyle 0.25}$ & $0.38_{\scriptstyle 0.23}$ & $1.00_{\scriptstyle 0.00}$ \\ 
				& MCP & 0.01 & $0.62_{\scriptstyle 0.11}$ & $0.97_{\scriptstyle 0.08}$ & $1.00_{\scriptstyle 0.00}$ & 0.04 & $0.69_{\scriptstyle 0.18}$ & $0.73_{\scriptstyle 0.21}$ & $1.00_{\scriptstyle 0.00}$ & 0.00 & $0.46_{\scriptstyle 0.16}$ & $0.63_{\scriptstyle 0.22}$ & $1.00_{\scriptstyle 0.00}$ & 0.00 & $0.38_{\scriptstyle 0.16}$ & $0.40_{\scriptstyle 0.16}$ & $1.00_{\scriptstyle 0.00}$ & 0.00 & $0.32_{\scriptstyle 0.15}$ & $0.22_{\scriptstyle 0.10}$ & $1.00_{\scriptstyle 0.00}$ \\ 
				\hline 
	\end{tabular}}}
\end{table}

\begin{table}[H]
	\centering
	\caption{Prop is the proportion of times true model being selected; for SCAD and MCP it corresponds to true model+intercept. Means and standard deviations(in subscript) for MCC, TPR and TNR over different methods are tabulated corresponding to $s_0 = 5$, $p = 5000$ and $n = 200$. ``AdLa" is short for Adaptive LASSO}\label{tab-p5000-n200}
	{\resizebox{\textwidth}{!}{\begin{tabular}{|c c||c c c c|c c c c|c c c c|c c c c|c c c c|}\hline
				\multirow{2}{*}{\parbox{0.07 \linewidth}{$p$=5000 \ $n$=200}} &  & \multicolumn{4}{|c|}{Independent} &  \multicolumn{4}{|c|}{Compound symmetry} & \multicolumn{4}{|c|}{\centering AR(1) with $\rho = 0.5$} & \multicolumn{4}{|c|}{\centering AR(1) with $\rho = 0.7$} & \multicolumn{4}{|c|}{\centering AR(1) with $\rho = 0.9$} \\ 
				& & Prop & MCC & TPR & TNR & Prop & MCC & TPR & TNR & Prop & MCC & TPR & TNR & Prop & MCC & TPR & TNR & Prop & MCC & TPR & TNR \\ 
				\hline 
				\multirow{4}{*}{\parbox{0.07 \linewidth}{\centering Set-1}} & SAVS & 0.96 & $1.00_{\scriptstyle 0.02}$ & $1.00_{\scriptstyle 0.00}$ & $1.00_{\scriptstyle 0.00}$ & 0.95 & $1.00_{\scriptstyle 0.03}$ & $1.00_{\scriptstyle 0.00}$ & $1.00_{\scriptstyle 0.00}$ & 0.97 & $1.00_{\scriptstyle 0.01}$ & $1.00_{\scriptstyle 0.00}$ & $1.00_{\scriptstyle 0.00}$ & 0.96 & $1.00_{\scriptstyle 0.03}$ & $0.99_{\scriptstyle 0.03}$ & $1.00_{\scriptstyle 0.00}$ & 0.28 & $0.82_{\scriptstyle 0.15}$ & $0.7_{\scriptstyle 0.23}$ & $1.00_{\scriptstyle 0.00}$ \\ 
				& S5 & 1.00 & $1.00_{\scriptstyle 0.00}$ & $1.00_{\scriptstyle 0.00}$ & $1.00_{\scriptstyle 0.00}$ & 1.00 & $1.00_{\scriptstyle 0.01}$ & $1.00_{\scriptstyle 0.01}$ & $1.00_{\scriptstyle 0.00}$ & 1.00 & $1.00_{\scriptstyle 0.00}$ & $1.00_{\scriptstyle 0.00}$ & $1.00_{\scriptstyle 0.00}$ & 0.99 & $1.00_{\scriptstyle 0.02}$ & $1.00_{\scriptstyle 0.04}$ & $1.00_{\scriptstyle 0.00}$ & 0.00 & $0.65_{\scriptstyle 0.15}$ & $0.46_{\scriptstyle 0.15}$ & $1.00_{\scriptstyle 0.00}$ \\
				& SCAD & 0.35 & $0.76_{\scriptstyle 0.25}$ & $1.00_{\scriptstyle 0.00}$ & $1.00_{\scriptstyle 0.00}$ & 0.62 & $0.88_{\scriptstyle 0.19}$ & $1.00_{\scriptstyle 0.00}$ & $1.00_{\scriptstyle 0.00}$ & 0.31 & $0.75_{\scriptstyle 0.24}$ & $1.00_{\scriptstyle 0.00}$ & $1.00_{\scriptstyle 0.00}$ & 0.06 & $0.53_{\scriptstyle 0.22}$ & $0.98_{\scriptstyle 0.07}$ & $1.00_{\scriptstyle 0.00}$ & 0.00 & $0.32_{\scriptstyle 0.18}$ & $0.43_{\scriptstyle 0.15}$ & $1.00_{\scriptstyle 0.00}$ \\ 
				& AdLa & 0.74 & $0.96_{\scriptstyle 0.08}$ & $1.00_{\scriptstyle 0.00}$ & $1.00_{\scriptstyle 0.00}$ & 0.59 & $0.94_{\scriptstyle 0.01}$ & $1.00_{\scriptstyle 0.00}$ & $1.00_{\scriptstyle 0.00}$ & 0.59 & $0.93_{\scriptstyle 0.10}$ & $1.00_{\scriptstyle 0.02}$ & $1.00_{\scriptstyle 0.00}$ & 0.27 & $0.83_{\scriptstyle 0.16}$ & $0.88_{\scriptstyle 0.180}$ & $1.00_{\scriptstyle 0.00}$ & 0.09 & $0.66_{\scriptstyle 0.21}$ & $0.55_{\scriptstyle 0.25}$ & $1.00_{\scriptstyle 0.00}$ \\ 
				& MCP & 0.62 & $0.88_{\scriptstyle 0.18}$ & $1.00_{\scriptstyle 0.00}$ & $1.00_{\scriptstyle 0.00}$ & 0.76 & $0.96_{\scriptstyle 0.08}$ & $1.00_{\scriptstyle 0.00}$ & $1.00_{\scriptstyle 0.00}$ & 0.60 & $0.88_{\scriptstyle 0.18}$ & $1.00_{\scriptstyle 0.00}$ & $1.00_{\scriptstyle 0.00}$ & 0.32 & $0.77_{\scriptstyle 0.21}$ & $0.98_{\scriptstyle 0.09}$ & $1.00_{\scriptstyle 0.00}$ & 0.00 & $0.43_{\scriptstyle 0.19}$ & $0.41_{\scriptstyle 0.16}$ & $1.00_{\scriptstyle 0.00}$ \\ 
				\hline 
				\multirow{4}{*}{\parbox{0.07 \linewidth}{\centering Set-2}} & SAVS & 0.94 & $0.99_{\scriptstyle 0.03}$ & $0.99_{\scriptstyle 0.03}$ & $1.00_{\scriptstyle 0.00}$ & 0.62 & $0.95_{\scriptstyle 0.07}$ & $0.92_{\scriptstyle 0.12}$ & $1.00_{\scriptstyle 0.00}$ & 0.81 & $0.97_{\scriptstyle 0.06}$ & $0.96_{\scriptstyle 0.1}$ & $1.00_{\scriptstyle 0.00}$ & 0.30 & $0.86_{\scriptstyle 0.12}$ & $0.77_{\scriptstyle 0.2}$ & $1.00_{\scriptstyle 0.00}$ & 0.00 & $0.62_{\scriptstyle 0.16}$ & $0.42_{\scriptstyle 0.18}$ & $1.00_{\scriptstyle 0.00}$ \\ 
				& S5 & 0.95 & $0.99_{\scriptstyle 0.02}$ & $0.99_{\scriptstyle 0.04}$ & $1.00_{\scriptstyle 0.00}$ & 0.59 & $0.95_{\scriptstyle 0.07}$ & $0.91_{\scriptstyle 0.12}$ & $1.00_{\scriptstyle 0.00}$ & 0.80 & $0.97_{\scriptstyle 0.07}$ & $0.94_{\scriptstyle 0.12}$ & $1.00_{\scriptstyle 0.00}$ & 0.19 & $0.84_{\scriptstyle 0.12}$ & $0.72_{\scriptstyle 0.18}$ & $1.00_{\scriptstyle 0.00}$ & 0.00 & $0.52_{\scriptstyle 0.18}$ & $0.35_{\scriptstyle 0.13}$ & $1.00_{\scriptstyle 0.00}$ \\
				& SCAD & 0.00 & $0.43_{\scriptstyle 0.14}$ & $1.00_{\scriptstyle 0.01}$ & $0.99_{\scriptstyle 0.00}$ & 0.01 & $0.56_{\scriptstyle 0.14}$ & $0.99_{\scriptstyle 0.04}$ & $1.00_{\scriptstyle 0.00}$ & 0.00 & $0.36_{\scriptstyle 0.13}$ & $0.98_{\scriptstyle 0.07}$ & $0.99_{\scriptstyle 0.00}$ & 0.00 & $0.31_{\scriptstyle 0.13}$ & $0.73_{\scriptstyle 0.20}$ & $0.99_{\scriptstyle 0.00}$ & 0.00 & $0.28_{\scriptstyle 0.17}$ & $0.33_{\scriptstyle 0.13}$ & $1.00_{\scriptstyle 0.00}$ \\ 
				& AdLa & 0.46 & $0.91_{\scriptstyle 0.11}$ & $0.99_{\scriptstyle 0.02}$ & $1.00_{\scriptstyle 0.00}$ & 0.18 & $0.82_{\scriptstyle 0.13}$ & $0.96_{\scriptstyle 0.08}$ & $1.00_{\scriptstyle 0.00}$ & 0.24 & $0.82_{\scriptstyle 0.16}$ & $0.89_{\scriptstyle 0.17}$ & $0.99_{\scriptstyle 0.00}$ & 0.09 & $0.70_{\scriptstyle 0.20}$ & $0.71_{\scriptstyle 0.23}$ & $1.00_{\scriptstyle 0.00}$ & 0.04 & $0.59_{\scriptstyle 0.22}$ & $0.46_{\scriptstyle 0.24}$ & $1.00_{\scriptstyle 0.00}$ \\ 
				& MCP & 0.16 & $0.74_{\scriptstyle 0.18}$ & $1.00_{\scriptstyle 0.01}$ & $1.00_{\scriptstyle 0.00}$ & 0.29 & $0.87_{\scriptstyle 0.11}$ & $0.97_{\scriptstyle 0.07}$ & $1.00_{\scriptstyle 0.00}$ & 0.04 & $0.62_{\scriptstyle 0.17}$ & $0.97_{\scriptstyle 0.10}$ & $1.00_{\scriptstyle 0.00}$ & 0.00 & $0.48_{\scriptstyle 0.16}$ & $0.67_{\scriptstyle 0.19}$ & $1.00_{\scriptstyle 0.00}$ & 0.00 & $0.35_{\scriptstyle 0.15}$ & $0.30_{\scriptstyle 0.13}$ & $1.00_{\scriptstyle 0.00}$ \\ 
				\hline 
	\end{tabular}}}
\end{table}

\bibliography{savsref}

\end{document}